\documentclass[aps,prb,preprint,showpacs,footinbib]{revtex4-1}

\usepackage{graphicx}
\usepackage{amsmath}
\usepackage{amsbsy}
\usepackage{braket}
\usepackage{amssymb}
\usepackage{dcolumn}
\usepackage{bm}
\usepackage[utf8]{inputenc}

\begin{document}

\title{Quantum Mechanical Assessment of Optimal Photovoltaic Conditions in Organic Solar Cells}

\author{Artur M. Andermann}
\address {Department of Physics, Universidade Federal de Santa Catarina, 88040-900, Florianópolis, Santa Catarina, Brazil}
\author{Luis G. C. Rego}
\address{Department of Physics, Universidade Federal de Santa Catarina, 88040-900, Florianópolis, Santa Catarina, Brazil}
\email{luis.guilherme@ufsc.br}

\begin{abstract}
Recombination losses contribute to reduce $J_{SC}$, $V_{OC}$ and the fill factor of  organic solar cells. 
Recent advances in non-fullerene organic photovoltaics have shown, nonetheless, that efficient charge generation can occur under small energetic driving forces 
($\Delta E_{DA}$) and low recombination losses.
To shed light on this issue, we set up a coarse-grained open quantum 
mechanical model for investigating the charge generation dynamics subject to various energy loss mechanisms. 
The influence of energetic driving force, Coulomb interaction, vibrational disorder, geminate recombination, temperature and external bias are included in the
analysis of the optimal photovoltaic conditions for charge carrier generation.
The assessment reveals that 
the overall energy losses are not only minimized when $\Delta E_{DA}$ approaches the effective reorganization energy at the interface
but also become insensitive to temperature and electric field variations.
It is also observed that a moderate reverse bias reduces geminate recombination losses significantly at vanishing driving forces,
where the charge generation is strongly affected by temperature.
\vspace{1.5cm}
\begin{center}
  \includegraphics[width=10cm]{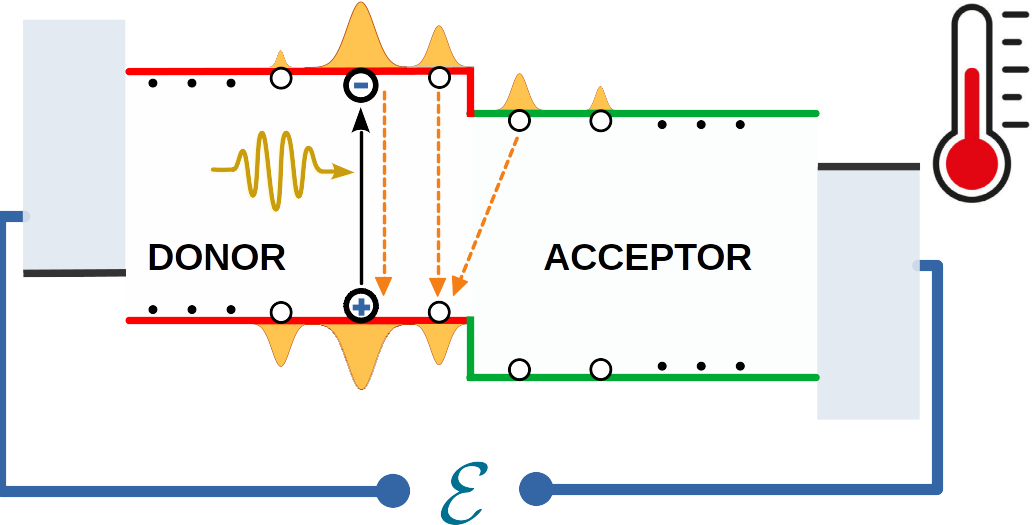}
\end{center}
\end{abstract}

\maketitle
\newpage

\vskip2pc


Efficient charge generation has always been a major concern for organic photovoltaics (OPV)
due to  the strong exciton binding energy in organic materials.\cite{Durrant2010,Armin2021,Cheng2018,hallermann2008charge,menke2014exciton} 
Thus, fullerene derivatives have been used as prototypical electron acceptor materials, because of their remarkable electron-accepting capabilities. 
However, the efficiency of fullerene-based organic solar cells (OCS)  has stagnated around 13\% 
as a consequence  of the excessive energy losses undergone to achieve  electron-hole (e-h) charge
separation, and the very poor optical characteristics of the C$_{60}$ molecule that does not contribute to charge generation.
Meanwhile, a new class of non-fullerene acceptor (NFA) materials have been developed, which revealed a new paradigm for OPV, one of efficient charge generation 
and high output voltage despite lower energetic gradients at the heterojunction.
Nowadays, the conversion efficiency of the NFA devices is approaching the barrier of 20\%,\cite{Armin2021} as a result of 
the high photo-absorption of non-fullerene small molecules; the push-pull design of the NFA molecules that facilitate the electron-hole pair separation 
upon photoexcitation; the complementary light-harvesting roles played by large band-gap donor materials and  low-bandgap NFA materials; 
besides the beneficial morphological properties of the NFA molecules ({\it e.g.}, miscibility, stability, and planarity).
For OSCs the output voltage is given by $V_{out} = E_{CT}/q -\Delta V_{rec}^{r} - \Delta V_{rec}^{nr}$,\cite{Rosenthal,Liu2016,Hou2018}
where $E_{CT}/q$ is related to the energetic driving force\cite{Liu2016,Andrienko} and the voltage losses due to recombination comprise both radiative 
($\Delta V_{rec}^{r}$) and non-radiative ($\Delta V_{rec}^{nr}$) pathways.
In NFA-OSCs the voltage loss mechanisms have been empirically reduced whereas the interplay among the relevant physical 
processes at the D:A interface remains a subject  of
study.\cite{Rosenthal,Liu2016,Andrienko,Bredas2019,Gao2015,JennyNelsonJACS,classen2020role,Chen2021,Recombination-temperature,Benduhn2017,Qian2018,Hofinger,Chen2018}

In this paper we  combine  the Ehrenfest and Redfield methods in a coarse-grained open quantum mechanical model to investigate, from a fundamental point of view,
the energy loss effects that occur during the free charge carrier generation process at the D:A interface. 
We study the influence of the energetic driving, temperature and external bias on the charge generation process.
The simulation results demonstrate that the overall energy losses are minimized at the activationless regime
of the charge separation, where driving force matches the effective reorganization energy of the heterojunction,
corroborating experimental studies.\cite{Coffey,Ward,Forrest}
Moreover, the optimal photovoltaic condition is insensitive to variations of temperature and reverse bias.


We partition the total Hamiltonian, responsible for the charge generation and recombination processes, as
\begin{equation}
H = H_{S} + H_{B} + H_{SB}~.
\end{equation} 
The system Hamiltonian, $H_S$, comprises the degrees of freedom of the electron and the hole and a subset of vibrational reorganization modes 
that are described as classical  modes within the framework of the  Ehrenfest self-consistent  method.
Thus, $H_S$ is  a mixed quantum-classical Hamiltonian (details provided in Support Information).
The bath Hamiltonian, $H_B$, accounts for  the  degrees of freedom of the environment, 
namely  the vibrational degrees of freedom and  fluctuations of the dielectric background, which we describe as an ensemble of quantum harmonic oscillators. 
The system-bath coupling ($H_{SB}$) is described within the framework of the  Redfield theory. 
Using the adiabatic representation for the system Hamiltonian, $H_S |\varphi_a\rangle = E_a |\varphi_a\rangle$, the Redfield equation for the reduced density matrix $\sigma$ reads
\begin{equation}\label{EQT1}
\frac{\partial \sigma_{ab}}{\partial t} = - i \omega_{ab} \sigma_{ab} + \sum_{c, d} R_{abcd} \sigma_{cd}
\end{equation}
where $\omega_{ab}=(E_a-E_b)/\hbar$ designates the eigenfrequencies of the electronic system 
and $R_{abcd}$ is the Redfield relaxation tensor.\cite{redfield1957theory, redfield1965theory} 
The first term on the right-hand-side (RHS) of Eq. \eqref{EQT1} describes the coherent quantum dynamics of S, the second term describes the interaction of 
S with the environment.
The implementation of the Redfield equations for this model has been described elsewhere.\cite{Artur}
Later on we will incorporate a term for the recombination effects into  Eq. \eqref{EQT1}.

\begin{figure}
    \centering
  \includegraphics[width = 0.7\linewidth]{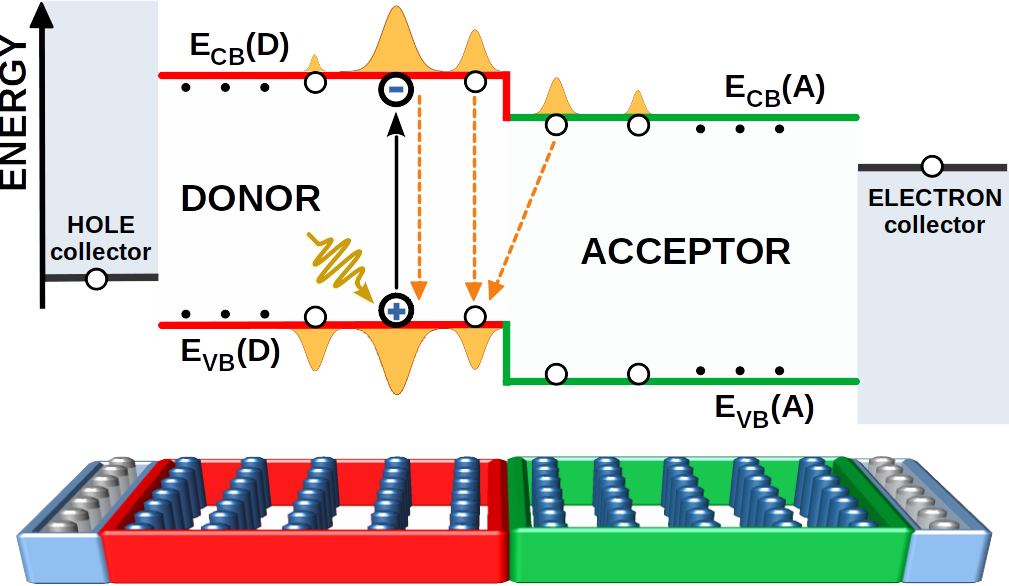}
  \caption{Model for the donor-acceptor interface. The figure on top shows the energetics of the D:A interface, including the electron and hole collectors. 
The gaussian envelopes depict the electron-hole excitation dynamics and charge separation. The upward arrow indicates the photoexcitation of the electron-hole pair 
(black) whereas the downward dashed arrows describe overall radiative and non-radiative recombination decays. At the bottom, the spatial configuration of 
the two-dimensional model of the D:A interface. Each molecule is described as a site in a 2D lattice.} 
 \label{model-pic}  
\end{figure}


The D:A interface is modelled as a two-dimensional (2D) coarse-grained lattice, with energy profile corresponding to a staggered (type II) interface, as described in 
Figure \ref{model-pic}.
The spatial arrangement of the  2D-D:A interface is  shown at the bottom of the figure, where each site of the lattice represents a molecular site in the 
coarse-grained model.  Periodic boundary conditions are applied along the transverse direction.
In the simulations, we consider the photoexcitation of a molecular site of the donor material, but the obtained results are equally valid for a 
photoexcitation in the acceptor material. 
After charge separation, 
the electron and hole are collected at the respective collecting layers at the border of the 2D lattice  (see Figure \ref{model-pic}).
The electron and hole collecting layers in the model do not represent the actual cathode and anode terminals of the device, they
are simply used as a theoretical tool to avoid the reflection of the wavepackets at the left and right boundaries;
they are positioned below $E_{CB}(A)$ and above $E_{VB}(D)$, respectively.

The coarse-grained Hamiltonian $H_S$, for either  the electron and hole photoexcited particles, is
\begin{equation}
H_S \equiv H^{\text{el/hl}} = \sum_{i}^{N}  \left\{E_i^{\text{el/hl}} + \varepsilon_i(t) - \Phi^{\text{el/hl}}_i(t)\right\} |i \rangle \langle i |  
+ \sum_{i \neq j}^{N} V_{ij}(t) |i \rangle \langle j |~,
\label{Hs}
\end{equation} 
where $N$ denotes the total number of molecular sites in the lattice.
The orthogonal basis set \{$|i\rangle\}$ consists of diabatic states associated with each of the molecular sites.
In addition, $E^{\textbf{el}}_i = E^{CB}_i$ and $E^{\textbf{hl}}_i = E^{VB}_i$ are the on-site energies of the electron (conduction band) and hole (valence band); 
$\varepsilon_i$ is the confinement energy associated with site $i$; 
and $\Phi^{\text{el/hl}}_i$ describes the electron-hole electrostatic interaction, given in the time-dependent mean-field  approximation by
\begin{equation} 
\Phi^{\text{el/hl}}_i(t)  = \xi_{bind}~ \left(\sum_j \frac{P^{\text{hl/el}}_j}{(1 + d_{ij})}\right)~. 
\label{bind} 
\end{equation} 
where $\xi_{bind}$ is the on-site electron-hole binding energy (refer to Table \ref{tbl:parameters} for a complete list of model parameters),  
$P^{\text{el}}_j$ ($P^{\text{hl}}_j$) is  the time-dependent electron (hole) population on site $j$ and $d_{ij}=|\vec{R}_i-\vec{R}_j|$ is the distance between 
molecular sites $i$  
and $j$.  
The treatment accounts for the decrease of the el-hl Coulomb barrier induced by the charge delocalization, besides describing the influence of temperature
in the el-hl separation, as shown elsewhere.\cite{Artur}
The last term in Eq. \eqref{Hs}, 
\begin{equation} 
V_{ij}  = V_0~F_{ij} =  
V_0 \left[
 \frac{ 2 \ell_{i}\ell_{j} }{ \ell_{i}^2+\ell_{j}^2} \exp \Big ({ \frac{- d^2_{ij}}{\ell_{i}^2+\ell_{j}^2} } \Big ) 
\right]~,
\label{hopping}
\end{equation}
describes the tunneling energy between lattice sites, where
$V_0$ is the bare electronic coupling and the form factor $F_{ij}$ is the time-dependent overlap between 
lattice sites $i$ and $j$, with $\ell$ being the confinement radius of the molecular site (see SI for details).
The time dependence of the energy parameters comprising Eq \eqref{Hs}, namely $\varepsilon_i(t)$ and $V_{ij}(t)$ gives rise to the 
{\it intra}-molecular and {\it inter}-molecular vibrational effects, respectively, that we associate with the Holstein and Peierls couplings.\cite{Artur}
The treatment includes dynamic disorder, which is taken into account by the intra-molecular
and inter-molecular electron-phonon couplings in the system Hamiltonian ($H_S$). The dynamic disorder is also
taken into account in the system-bath Hamiltonian ($H_{SB}$) that gives rise to the Redfield relaxation tensor and,
lastly, through the action of the classical Berendsen thermostat that acts on the molecular sites (describe in SI). 
These interactions give rise to localization of the electron and hole wavepackets,\cite{candiotto2017charge} polaron formation,\cite{Artur} and
temperature dependence of the charge generation rate. The model does not include static disorder. 


Recombination is the main efficiency loss mechanism in photovoltaics, even more so in OSCs due to the convoluted charge separation process 
in organic devices.
Recombination processes can be separated in two types: geminate and non-geminate (including bimolecular and trap-assisted processes). 
The former designates the annihilation of a correlated
electron-hole pair generated by the same photoexcitation event, or the recombination of a charge separated (CT) state before complete dissociation. 
Experiments have shown that geminate recombination is more important at short-circuit and low illumination conditions.\cite{Durrant2012}
Bimolecular recombination, on the other hand, involves fully dissociated carriers generated by independent photoabsorption  events. It is more relevant
at open-circuit and/or high excitation  conditions. 
A quantum mechanical model that is well suited for describing the geminate recombination in molecular systems is the Haberkorn model. 
It was initially proposed to describe recombination in radical pairs,\cite{Haberkorn} 
but later  was used to describe the trapping and exciton recombination 
in the Fenna-Matthews-Olson (FMO) protein complex\cite{Aspuru} and OSC models.\cite{candiotto2017charge} 
Herein we include the vibrational dynamics, in addition to the electron-hole dynamics
to extend the plain Haberkorn formalism.  Thus, consider a quantum state
$|\Psi^\eta(t)\rangle = \sum_i C^\eta_i(t) |i\rangle$ that comprises the reduced density matrix $\sigma^\eta$,
where $\eta = el,hl$ designates the type of charge carrier. As in the Haberkorn model, the mutual annihilation of electrons and holes is given by the 
dynamical equation
\begin{equation}
\frac{d}{dt}|\Psi^\eta(t)\rangle = -\hat{\Gamma}^{\eta'}_R|\Psi^\eta(t)\rangle
\label{rec-dyn}
\end{equation}
where $\eta'$ is the carrier type opposite to $\eta$. To include the vibrational effects into the Haberkorn recombination model, we use the 
 operator $\hat{\Gamma}^{\eta'}_R$, defined  as
\begin{equation}
\hat{\Gamma}^{\eta'}_R = |\Psi^{\eta'}(t)\rangle~\kappa\hat{F}~\langle \Psi^{\eta'}(t)|~,
\label{rec-oper}
\end{equation}
where the symmetric operator $\kappa \hat{F} = \sum_{i} \kappa F_{ii} |i\rangle\langle i| + \sum_{i\neq j} \kappa F_{ij} |i\rangle\langle j|$ 
includes the molecular site vibrations, as defined in Eq. \eqref{hopping}.
The parameter $\kappa$ in Eq. (\ref{rec-oper}) (see Table \ref{tbl:parameters}) is a molecular decay rate constant. 
Coropceanu and collaborators\cite{Bredas2019,Chen2021} associate $\kappa F_{ii}$ with the Einstein spontaneous emission rate for the radiative 
recombination term whereas the Marcus-Levich-Jortner (MLJ) formula was used to define the non-radiative decay rate $\kappa F_{ij}$;
herein we use the same parameter $\kappa$ for both processes. A comparison between the vibronic Haberkorn model and the three-state 
model\cite{Bredas2019,Benduhn2017,Qian2018,JennyNelsonJACS,Chen2021} is presented as SI.
Substitution of Eq. (\ref{rec-oper}) into Eq. (\ref{rec-dyn}) yields the following equation for the coefficients of the wavepacket in the molecular site basis
\begin{equation}
\frac{d}{dt}C^\eta_i = - \kappa\left(\sum_{k,l} \left[C^{\eta'}_k\right]^*F_{kl}~C^\eta_l\right) C^{\eta'}_i~.
\label{C-eta-dot}
\end{equation}
Rewriting Eq. (\ref{C-eta-dot}) in terms of the time derivative of the reduced density matrix, $\sigma^\eta_{ij} = C^\eta_i \left[C^{\eta}_j\right]^*$, we get
\begin{equation}
\frac{d}{dt} \sigma^\eta\Big|_{rec} = - \kappa \left( \sigma^{\eta'} \hat{F} \sigma^\eta + \sigma^\eta \hat{F}\sigma^{\eta'} \right)~.
\label{master-rec}
\end{equation}
The quantum mechanical recombination rate given by Eq. (\ref{master-rec}), for  geminate recombination, can be associated with the classical law of mass action for the rate of 
bimolecular recombination $R=knp$, where $n$ and $p$ are the concentrations of free electrons and holes, respectively, and $k$ is an effective 
recombination constant given, for instance, by the Langevin model.\cite{Burke}
It is also worth noting that Eq. \eqref{master-rec} gives the recombination losses during the exciton diffusion and for the
CT state at the D:A interface.

Before adding this recombination rate term to the Redfield Master Equation, Eq. \eqref{EQT1},
it is transformed to the adiabatic representation of basis states $\{|\varphi_a\rangle\}$ by 
$\dot{\sigma}^{\eta}_{rec} \rightarrow  (\hat{U}^{\eta})^{-1}~\dot{\sigma}^{\eta}_{rec}~\hat{U}^{\eta}$, 
where $\hat{U}^{\eta}$ is the unitary operator that diagonalizes $\hat{H}^{\eta}$. 
The overall charge neutrality of the heterojunction is conserved with Eq. \eqref{master-rec}. 
Based on Eq. (\ref{master-rec}), the energy lost by a molecular site via recombination is calculated as 
$W^{rec}_i = \sigma^{rec}_{ii}(t)E^{gap}_i$
where $\sigma^{rec}_{ii}$ is the population lost by recombination and $E^{gap}_i$ is the corresponding CB-VB gap. The total energy loss due to recombination
is calculated as $\sum_i W^{rec}_i(t_f)$, where the sum runs over all the molecular sites and $t_f$ is the final simulation time. 
Therefore, to obtain the  voltage loss due to geminate recombination ($\Delta V^{rec}$) during the charge separation process we calculate
\begin{equation}
\Delta V^{rec} = \frac{1}{e}\sum_i W^{rec}_i(t_f)
\label{DV}
\end{equation}
where $e$ is the electronic charge. Notice that this method  accounts only for the geminate electron-hole recombination. Bimolecular 
recombination processes become important at open circuit conditions and  high excitation intensities, which are not described by the present simulations.

The model parameters used in the simulations are presented in Table \ref{tbl:parameters}.
The set of coupled differential equations produced by combining Eqs. \eqref{EQT1} and \eqref{master-rec} are solved numerically by a 
fourth order adaptive step size Runge-Kutta method.
We simulate the  process of free charge generation, including exciton photoexcitation, excition diffusion, CT state formation and separation, 
for a single el-hl pair in the presence of geminate recombination, during a  time frame of ten picoseconds.

\begin{table}
    \caption{Values and description of the parameters used in calculations (refer to SI for details).}
    \label{tbl:parameters}
    \begin{tabular}{ccc}
\hline
        & value  & description \\
\hline
         $N$                & 84            & number of sites \\
        
         $\ell_{\text{o}}$ & 0.5 nm        & confinement radius for the empty molecular site \\
        
         $V^{nn}$ & $-40$ meV & nearest-neighbor electronic coupling \\
        
          $d_{ij}$  & 1.3 nm   & distance of nearest neighbor sites \\
        
         $\xi_{\text{bind}}$ & 0.2 eV   & electron-hole coupling constant \\
        
         $\lambda$ & 50 meV & reorganization energy \\
        
         $\omega_c$  & 1334 $\text{cm}^{-1}$ (165 meV) & cutoff frequency \\
        
         $\Omega$ &  272 $\text{cm}^{-1}$ (34 meV) & classical normal mode frequency \\
        
        $\gamma^{-1}$ & $0.1$ ps & phonon-phonon relaxation time \\
		$\kappa$        & 1 ps$^{-1}$   & molecular site decay rate constant \\
\hline
        
    \end{tabular}
\end{table}

We consider that a Frenkel exciton  is photoexcited in the center of the donor region, as shown in Figure \ref{model-pic}.
The coarse-grained model consists of a 2D-square lattice with nearest neighbor (nn) distance $a$ = 1.3 nm and  equal D and A domain
lengths of approximately 7.5 nm. The effective electronic
coupling between nearest neighbors is set to be $V^{nn} = -40$ meV, as given by Eq. (\ref{hopping}), which is in agreement with other molecular crystal
models.\cite{lee2015charge, coropceanu2007charge, mozafari2013polaron, kocherzhenko2015coherent, de2017vibronic, troisi2006dynamics,candiotto2017charge}
The low frequency vibration mode $\Omega$ is treated classically. It modulates the radii $\ell_i(t)$ of the molecular sites and, thereby, the electronic 
coupling $V_{ij}$, as given by \eqref{hopping}. The classical vibrations are self-consistently coupled to the el-hl quantum states by semiclassical 
Ehrenfest forces (Supporting Information).  
The  high energy quantum vibrations are incorporated into the Redfield relaxation tensor, up to the cutoff frequency $\omega_c$ = 1334
cm$^{-1}$ that corresponds to the localized C$=$C stretching mode.
For the system-bath coupling we use a reorganization energy parameter of $\lambda$ = 50 meV for either the electron and hole density matrices.
The magnitude of $\lambda$ is commonly found to be in the range of 100-300 meV for organic blends.\cite{Vandewal2017}
However, it has been shown that the values of $\lambda$ obtained from experiments tend to be larger than the bare values obtained by static calculations.
Vanderwal et al.\cite{Vandewal2017} derived the empirical
relationship $\lambda_{exp} = \lambda_{calc} + \Delta$, where $\Delta \approx$ 95 meV is a correction due to dynamic
disorder (disregarding the static disorder). Additionally, McGehee et al.\cite{Burke} derived another relation in which 
$\lambda_{exp} = \lambda_{calc} + \sigma^2/(2kT)$, where $\sigma$ is the static energetic disorder parameter and $\sigma^2/(2kT)\approx$  108 meV
at room temperature, assuming $\sigma$ = 75 meV. $\lambda_{exp}$ can be associated with the CT emission spectral broadening that is caused by disorder.
In our model we ascribe the effects of disorder fully to the molecular vibrations.
Therefore, the effective reorganization energy affecting the el-hl  dynamics in our simulations
can be estimated as $\lambda_{eff} \approx 50 meV + 100 meV = 150 meV$. This is in good agreement
with the optimal charge generation conditions obtained in our simulations, which is given
$E_{DA} \approx \lambda_{eff}$ that occurs between 100 and 200 meV, as shown ahead.
As for the recombination rate constant, $\kappa$, its value is defined based on the characteristics of the coarse-grained model.
Experiments have determined that the exciton diffusion length in organic semiconductors and polymers is within the range of 
5-20 nm,\cite{Exc-diffusion,Firdaus2020} and the diffusion length is limited by the exciton lifetime. 
In our coarse-grained model the length of the donor (D) domain  is about 7.5 nm, which should correspond approximately to an exciton lifetime. 
To follow this constraint and to evince the effects of the recombination within the span of the simulation time (10 ps), we use $\kappa$ = 1 ps$^{-1}$. 
For the sake of comparison, Figure S4 in the Supporting Information presents simulation results with and without recombination. 
Finally, in accordance with the model (Figure \ref{model-pic}), we define the energetic driving force of the D:A interface  as
$E_{DA}\equiv  E^{{CB}}(D) - E^{{VB}}(A)$. 


\begin{figure}[htb]
    \centering
  \includegraphics[width = 0.8\linewidth]{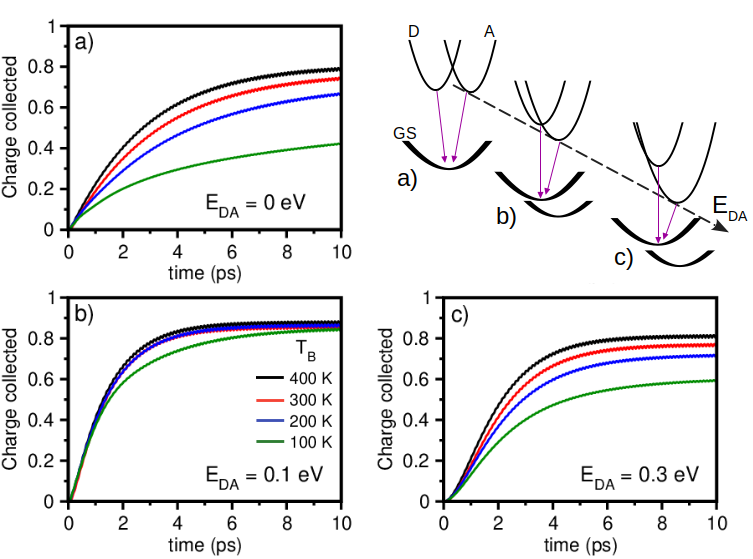}
  \caption{Collected electronic population ($N_e$) subjected to recombination losses, for various bath temperatures (T$_B$) and donor-acceptor energy offsets:
a) $E_{DA}$ = 0 eV, b) $E_{DA}$ = 0.1 eV and c) $E_{DA}$ = 0.3 eV.
The illustration at the top-right corner is a reference to the Marcus Theory, depicting electron transfer in the normal regime (a), the activationless regime (b), 
and the inverted regime (c) as the energetic driving force $E_{DA}$ increases.}
 \label{total-col-temperature} 
\end{figure}

The evaluation of the activation energy (E$_a$) required for  el-hl pair separation has been open to 
debate,\cite{Gao2015,Dong2019,Andrienko,Rosenthal,Coffey,Ward,Unger} 
given that the  obtained values may vary depending on the technique and the time scale of the measurements. From the theoretical viewpoint, 
considering the kinetic models, 
E$_a$ is an effective parameter that is influenced by Coulombic dissociation and recombination rates.\cite{Gao2015,Dong2019,Andrienko}
Figure \ref{total-col-temperature} shows the collected electronic population after recombination, for various bath temperatures  (T$_B$)
and  energetic driving forces ($E_{DA}$), that is $N_e(E_{DA},T_B)$, without electric field bias.
The simulations show, as a whole,  that the charge generation rate increases with temperature for all D:A offsets.
In addition, the effect of T$_B$ on $N_e$  reveals three distinct electron separation regimes, as depicted in the inset of Figure \ref{total-col-temperature}:
a) the normal electron separation for negligible $E_{DA}$ = 0 eV; b) the activationless regime for $E_{DA} \approx$ 0.1 eV; and c) the inverted Marcus regime for 
$E_{DA}$ = 0.3 eV. The highest internal quantum efficiency (IQE) is obtained for
the activationless regime,  $E_{DA} \approx$ 0.1 eV, whereby the charge generation process is almost independent of the temperature.
The maximum IQE  is obtained for $E_{DA} \approx \lambda_{eff}$, where $\lambda_{eff}$ corresponds to the effective reorganization energy at the 
heterojunction, corroborating experimental studies that show the existence of such an optimal driving force.\cite{Coffey,Ward,Forrest}
However, for vanishing energetic driving force, $E_{DA} \approx$ 0, and no electric fields (to be discussed next), the charge generation is very 
sensitive to the temperature.
Lastly, if the energy offset becomes excessive, {\it i.e.} $E_{DA} \gg \lambda$, the charge generation process enters the inverted region described by the 
Marcus-Jortner-Levich theory {\cite{Jortner,MJL,VSBatista}}, where charge transfer is assisted by quantum vibronic effects. 
A detailed account of the impact produced by vibronic effects (such as polaron formation) and internal electric fields on the free charge carrier generation  
is was presented by Andermann and Rego elsewhere.\cite{Artur}
Figures S2 and S3 of the SI evince the importance of electron-hole interaction and vibrational dynamics to recombination, respectively.

\begin{figure}[htp]
    \centering
  \includegraphics[width = 1.0\linewidth]{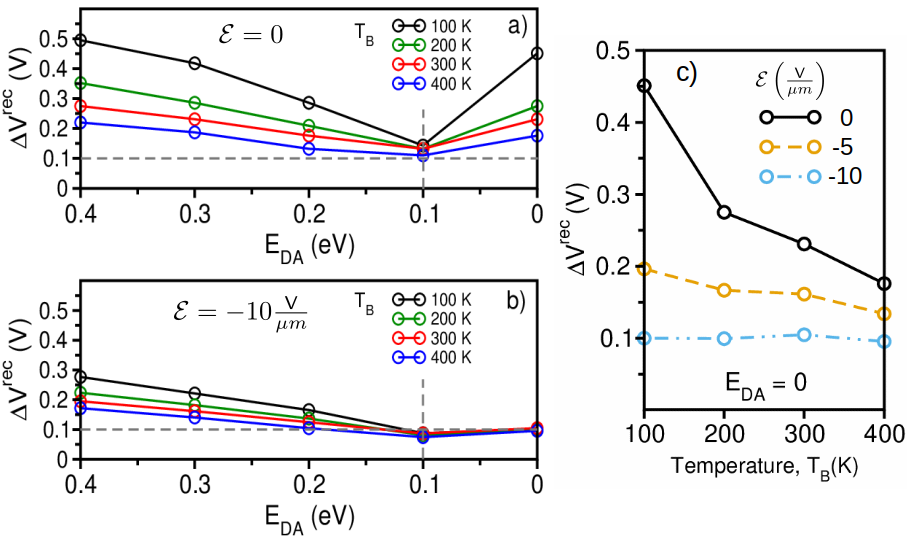}
  \caption{a) Geminate recombination induced voltage-loss,  $\Delta V^{rec}(E_{DA},T_B,\mathcal{E})$, as calculated with Eq. (\ref{DV}), 
without bias field ($\mathcal{E}$=0),  
for T$_B$ = 100, 200, 300 and 400 K.
b) Same as in a) with a reverse bias  field $\mathcal{E}$ =  -10 V/$\mu$m.
The dashed lines indicate the approximate position of the optimal charge generation regime, which does not change appreciably under the influence of
temperature or reverse bias. c) Temperature and bias field dependence of geminate recombination voltage-loss without D:A driving force, 
$\Delta V^{rec}(E_{DA}=0,T_B,\mathcal{E})$.}
 \label{delta_V} 
\end{figure}

In addition to the energetic driving force for charge separation, the voltage losses in the OSCs also include the contributions from recombination
processes, both radiative and non-radiative.
Figure \ref{delta_V} shows the effect of  driving force ($E_{DA}$),  temperature ($T_B$) and  electric field bias ($\mathcal{E}$) 
on the voltage loss due to geminate recombination, $\Delta V_{rec}(E_{DA},T_B,\mathcal{E})$, as given by Eq. \eqref{DV}.
The simulations show 
that $\Delta V_{rec}$ is practically insensitive to temperature and reverse bias field if the energetics of the heterojunction is set to the activationless
regime of charge separation, in this case $E_{DA}$ = 0.1 eV. Out of this point the losses due to geminate recombination at the D:A interface increase 
as the temperature decreases, particularly so for vanishing D:A offset in the  absence of reverse bias (Figure \ref{delta_V}-a). 
The influence of  temperature on the geminate recombination losses during charge carrier-generation has not been extensively investigated
because it is difficult to measure geminate recombination at low illumination intensities, since parasitic leakage currents can affect the 
results.\cite{Proctor-leakage} 
To study this issue, Gao et al.\cite{Gao2015} used low excitation intensities to observe 
a decreasing  V$_{OC}$ at low temperatures, which was ascribed to reduced carrier generation.
Here, it is important to distinguish the nonequilibrium carrier generation process that is observed at low illumination intensities from the 
quasi-equilibrium conditions attained at open-circuit with moderate-to-high illumination.  
In the first case, the efficiency of charge generation is determined by the competition between photogeneration of free charge carriers and geminate recombination,
as described in Figure \ref{delta_V}.
In the second case, bimolecular recombination kinetics is the main cause of energy loss, so that at the steady-state
$qV_{OC} = E_g -k_BT\ln(N_cN_v/np)$, where $n$ and $p$ are the electron and hole stationary free carrier densities.\cite{Burke,Gao2015,Rosenthal}
Thus, at high excitation densities $n$ and $p$ are nearly temperature independent and $V_{OC}$  decreases linearly with temperature.\cite{Gao2015,Rauh}
At low excitations, however, the density of free carriers become temperature dependent, $n(T)$ and $p(T)$,\cite{Gao2015,Rauh,Forrest,Recombination-temperature} 
thereby changing the behavior of $V_{OC}(T)$.
Chan et al.\cite{Recombination-temperature} reported that as the blends P3TEA:SF-PDI$_2$ and PM6:Y6 are cooled,
the reduced charge separation rate leads to increasing charge recombination, either directly at the D:A interface or via the LE state.
Thus, in this case $V_{OC}$ can exhibit a sublinear dependence on temperature,\cite{Inganas-2010,Forrest,Rauh} 
or even increase with temperature\cite{Gao2015} if the carrier-generation process dominates $V_{OC}$. 
Notice, though, that the application of a reverse bias at the heterojunction should promote the free carrier generation,\cite{Proctor} thereby decreasing the 
losses by geminate recombination and improving $V_{OC}$, as shown in Figure \ref{delta_V}-b.
The effect of a reverse bias  is particularly strong in the absence of driving force ($E_{DA}$=0), as shown in Figure \ref{delta_V}-c, where it is shown 
that a moderate bias reduces recombination losses substantially.

The various recombination pathways contribute to reduce $J_{SC}$, $V_{OC}$ and the fill factor (FF) of the OSCs, thereby limiting its 
power conversion efficiency. In regard to free carrier generation, this work indicates that the best device performances should be realized at 
the optimal offset $E_{DA} \approx \lambda_{eff}$, which produces low overall energy losses for free carrier generation and superior $J_{SC}$ and FF.

{\bf Acknowledgements.}
AMA is grateful for the financial support from FAPESC (Fundação de Amparo à Pesquisa do Estado de Santa Catarina). 
LGCR acknowledges support  by Coordena\c c\~ao de Aperfei\c coamento de Pessoal de N\'{\i}vel Superior Brasil (CAPES) - Finance Code 001, 
by the Brazilian National Counsel of Technological and Scientific Development (CNPq) and the National Institute for Organic Electronics (INEO).
The authors acknowledge support by allocation of supercomputer time from Laboratory for Scientific Computing (LNCC/MCTI, Brazil).

\bibliography{MyRefs}

\begin{thebibliography}{53}%
\makeatletter
\providecommand \@ifxundefined [1]{%
 \@ifx{#1\undefined}
}%
\providecommand \@ifnum [1]{%
 \ifnum #1\expandafter \@firstoftwo
 \else \expandafter \@secondoftwo
 \fi
}%
\providecommand \@ifx [1]{%
 \ifx #1\expandafter \@firstoftwo
 \else \expandafter \@secondoftwo
 \fi
}%
\providecommand \natexlab [1]{#1}%
\providecommand \enquote  [1]{``#1''}%
\providecommand \bibnamefont  [1]{#1}%
\providecommand \bibfnamefont [1]{#1}%
\providecommand \citenamefont [1]{#1}%
\providecommand \href@noop [0]{\@secondoftwo}%
\providecommand \href [0]{\begingroup \@sanitize@url \@href}%
\providecommand \@href[1]{\@@startlink{#1}\@@href}%
\providecommand \@@href[1]{\endgroup#1\@@endlink}%
\providecommand \@sanitize@url [0]{\catcode `\\12\catcode `\$12\catcode
  `\&12\catcode `\#12\catcode `\^12\catcode `\_12\catcode `\%12\relax}%
\providecommand \@@startlink[1]{}%
\providecommand \@@endlink[0]{}%
\providecommand \url  [0]{\begingroup\@sanitize@url \@url }%
\providecommand \@url [1]{\endgroup\@href {#1}{\urlprefix }}%
\providecommand \urlprefix  [0]{URL }%
\providecommand \Eprint [0]{\href }%
\providecommand \doibase [0]{http://dx.doi.org/}%
\providecommand \selectlanguage [0]{\@gobble}%
\providecommand \bibinfo  [0]{\@secondoftwo}%
\providecommand \bibfield  [0]{\@secondoftwo}%
\providecommand \translation [1]{[#1]}%
\providecommand \BibitemOpen [0]{}%
\providecommand \bibitemStop [0]{}%
\providecommand \bibitemNoStop [0]{.\EOS\space}%
\providecommand \EOS [0]{\spacefactor3000\relax}%
\providecommand \BibitemShut  [1]{\csname bibitem#1\endcsname}%
\let\auto@bib@innerbib\@empty
\bibitem [{\citenamefont {Clarke}\ and\ \citenamefont
  {Durrant}(2010)}]{Durrant2010}%
  \BibitemOpen
  \bibfield  {author} {\bibinfo {author} {\bibfnamefont {T.~M.}\ \bibnamefont
  {Clarke}}\ and\ \bibinfo {author} {\bibfnamefont {J.~R.}\ \bibnamefont
  {Durrant}},\ }\href {\doibase 10.1021/cr900271s} {\bibfield  {journal}
  {\bibinfo  {journal} {Chem. Rev.}\ }\textbf {\bibinfo {volume} {110}},\
  \bibinfo {pages} {6736} (\bibinfo {year} {2010})}\BibitemShut {NoStop}%
\bibitem [{\citenamefont {Armin}\ \emph {et~al.}(2021)\citenamefont {Armin},
  \citenamefont {Li}, \citenamefont {Sandberg}, \citenamefont {Xiao},
  \citenamefont {Ding}, \citenamefont {Nelson}, \citenamefont {Neher},
  \citenamefont {Vandewal}, \citenamefont {Shoaee}, \citenamefont {Wang},
  \citenamefont {Ade}, \citenamefont {Heumüller}, \citenamefont {Brabec},\
  and\ \citenamefont {Meredith}}]{Armin2021}%
  \BibitemOpen
  \bibfield  {author} {\bibinfo {author} {\bibfnamefont {A.}~\bibnamefont
  {Armin}}, \bibinfo {author} {\bibfnamefont {W.}~\bibnamefont {Li}}, \bibinfo
  {author} {\bibfnamefont {O.~J.}\ \bibnamefont {Sandberg}}, \bibinfo {author}
  {\bibfnamefont {Z.}~\bibnamefont {Xiao}}, \bibinfo {author} {\bibfnamefont
  {L.}~\bibnamefont {Ding}}, \bibinfo {author} {\bibfnamefont {J.}~\bibnamefont
  {Nelson}}, \bibinfo {author} {\bibfnamefont {D.}~\bibnamefont {Neher}},
  \bibinfo {author} {\bibfnamefont {K.}~\bibnamefont {Vandewal}}, \bibinfo
  {author} {\bibfnamefont {S.}~\bibnamefont {Shoaee}}, \bibinfo {author}
  {\bibfnamefont {T.}~\bibnamefont {Wang}}, \bibinfo {author} {\bibfnamefont
  {H.}~\bibnamefont {Ade}}, \bibinfo {author} {\bibfnamefont {T.}~\bibnamefont
  {Heumüller}}, \bibinfo {author} {\bibfnamefont {C.}~\bibnamefont {Brabec}},
  \ and\ \bibinfo {author} {\bibfnamefont {P.}~\bibnamefont {Meredith}},\
  }\href {\doibase https://doi.org/10.1002/aenm.202003570} {\bibfield
  {journal} {\bibinfo  {journal} {Adv. Energy Mater.}\ }\textbf {\bibinfo
  {volume} {11}},\ \bibinfo {pages} {2003570} (\bibinfo {year}
  {2021})}\BibitemShut {NoStop}%
\bibitem [{\citenamefont {Cheng}\ \emph {et~al.}(2018)\citenamefont {Cheng},
  \citenamefont {Li}, \citenamefont {Zhan},\ and\ \citenamefont
  {Yang}}]{Cheng2018}%
  \BibitemOpen
  \bibfield  {author} {\bibinfo {author} {\bibfnamefont {P.}~\bibnamefont
  {Cheng}}, \bibinfo {author} {\bibfnamefont {G.}~\bibnamefont {Li}}, \bibinfo
  {author} {\bibfnamefont {X.}~\bibnamefont {Zhan}}, \ and\ \bibinfo {author}
  {\bibfnamefont {Y.}~\bibnamefont {Yang}},\ }\href {\doibase
  10.1038/s41566-018-0104-9} {\bibfield  {journal} {\bibinfo  {journal} {Nat.
  Photonics}\ }\textbf {\bibinfo {volume} {12}},\ \bibinfo {pages} {131}
  (\bibinfo {year} {2018})}\BibitemShut {NoStop}%
\bibitem [{\citenamefont {Hallermann}\ \emph {et~al.}(2008)\citenamefont
  {Hallermann}, \citenamefont {Haneder},\ and\ \citenamefont
  {Da~Como}}]{hallermann2008charge}%
  \BibitemOpen
  \bibfield  {author} {\bibinfo {author} {\bibfnamefont {M.}~\bibnamefont
  {Hallermann}}, \bibinfo {author} {\bibfnamefont {S.}~\bibnamefont {Haneder}},
  \ and\ \bibinfo {author} {\bibfnamefont {E.}~\bibnamefont {Da~Como}},\
  }\href@noop {} {\bibfield  {journal} {\bibinfo  {journal} {Appl. Phys.
  Lett.}\ }\textbf {\bibinfo {volume} {93}},\ \bibinfo {pages} {290} (\bibinfo
  {year} {2008})}\BibitemShut {NoStop}%
\bibitem [{\citenamefont {Menke}\ and\ \citenamefont
  {Holmes}(2014)}]{menke2014exciton}%
  \BibitemOpen
  \bibfield  {author} {\bibinfo {author} {\bibfnamefont {S.~M.}\ \bibnamefont
  {Menke}}\ and\ \bibinfo {author} {\bibfnamefont {R.~J.}\ \bibnamefont
  {Holmes}},\ }\href@noop {} {\bibfield  {journal} {\bibinfo  {journal} {Energ.
  \& Environ. Sci.}\ }\textbf {\bibinfo {volume} {7}},\ \bibinfo {pages} {499}
  (\bibinfo {year} {2014})}\BibitemShut {NoStop}%
\bibitem [{\citenamefont {Rosenthal}\ \emph {et~al.}(2019)\citenamefont
  {Rosenthal}, \citenamefont {Hughes}, \citenamefont {Luginbuhl}, \citenamefont
  {Ran}, \citenamefont {Karki}, \citenamefont {Ko}, \citenamefont {Hu},
  \citenamefont {Wang}, \citenamefont {Ade},\ and\ \citenamefont
  {Nguyen}}]{Rosenthal}%
  \BibitemOpen
  \bibfield  {author} {\bibinfo {author} {\bibfnamefont {K.~D.}\ \bibnamefont
  {Rosenthal}}, \bibinfo {author} {\bibfnamefont {M.~P.}\ \bibnamefont
  {Hughes}}, \bibinfo {author} {\bibfnamefont {B.~R.}\ \bibnamefont
  {Luginbuhl}}, \bibinfo {author} {\bibfnamefont {N.~A.}\ \bibnamefont {Ran}},
  \bibinfo {author} {\bibfnamefont {A.}~\bibnamefont {Karki}}, \bibinfo
  {author} {\bibfnamefont {S.-J.}\ \bibnamefont {Ko}}, \bibinfo {author}
  {\bibfnamefont {H.}~\bibnamefont {Hu}}, \bibinfo {author} {\bibfnamefont
  {M.}~\bibnamefont {Wang}}, \bibinfo {author} {\bibfnamefont {H.}~\bibnamefont
  {Ade}}, \ and\ \bibinfo {author} {\bibfnamefont {T.-Q.}\ \bibnamefont
  {Nguyen}},\ }\href {\doibase https://doi.org/10.1002/aenm.201901077}
  {\bibfield  {journal} {\bibinfo  {journal} {Adv. Energy Mater.}\ }\textbf
  {\bibinfo {volume} {9}},\ \bibinfo {pages} {1901077} (\bibinfo {year}
  {2019})}\BibitemShut {NoStop}%
\bibitem [{\citenamefont {Liu}\ \emph {et~al.}(2016)\citenamefont {Liu},
  \citenamefont {Chen}, \citenamefont {Qian}, \citenamefont {Gautam},
  \citenamefont {Yang}, \citenamefont {Zhao}, \citenamefont {Bergqvist},
  \citenamefont {Zhang}, \citenamefont {Ma}, \citenamefont {Ade}, \citenamefont
  {Ingan{\"a}s}, \citenamefont {Gundogdu}, \citenamefont {Gao},\ and\
  \citenamefont {Yan}}]{Liu2016}%
  \BibitemOpen
  \bibfield  {author} {\bibinfo {author} {\bibfnamefont {J.}~\bibnamefont
  {Liu}}, \bibinfo {author} {\bibfnamefont {S.}~\bibnamefont {Chen}}, \bibinfo
  {author} {\bibfnamefont {D.}~\bibnamefont {Qian}}, \bibinfo {author}
  {\bibfnamefont {B.}~\bibnamefont {Gautam}}, \bibinfo {author} {\bibfnamefont
  {G.}~\bibnamefont {Yang}}, \bibinfo {author} {\bibfnamefont {J.}~\bibnamefont
  {Zhao}}, \bibinfo {author} {\bibfnamefont {J.}~\bibnamefont {Bergqvist}},
  \bibinfo {author} {\bibfnamefont {F.}~\bibnamefont {Zhang}}, \bibinfo
  {author} {\bibfnamefont {W.}~\bibnamefont {Ma}}, \bibinfo {author}
  {\bibfnamefont {H.}~\bibnamefont {Ade}}, \bibinfo {author} {\bibfnamefont
  {O.}~\bibnamefont {Ingan{\"a}s}}, \bibinfo {author} {\bibfnamefont
  {K.}~\bibnamefont {Gundogdu}}, \bibinfo {author} {\bibfnamefont
  {F.}~\bibnamefont {Gao}}, \ and\ \bibinfo {author} {\bibfnamefont
  {H.}~\bibnamefont {Yan}},\ }\href {\doibase 10.1038/nenergy.2016.89}
  {\bibfield  {journal} {\bibinfo  {journal} {Nat. Energy}\ }\textbf {\bibinfo
  {volume} {1}},\ \bibinfo {pages} {16089} (\bibinfo {year}
  {2016})}\BibitemShut {NoStop}%
\bibitem [{\citenamefont {Hou}\ \emph {et~al.}(2018)\citenamefont {Hou},
  \citenamefont {Ingan{\"a}s}, \citenamefont {Friend},\ and\ \citenamefont
  {Gao}}]{Hou2018}%
  \BibitemOpen
  \bibfield  {author} {\bibinfo {author} {\bibfnamefont {J.}~\bibnamefont
  {Hou}}, \bibinfo {author} {\bibfnamefont {O.}~\bibnamefont {Ingan{\"a}s}},
  \bibinfo {author} {\bibfnamefont {R.~H.}\ \bibnamefont {Friend}}, \ and\
  \bibinfo {author} {\bibfnamefont {F.}~\bibnamefont {Gao}},\ }\href {\doibase
  10.1038/nmat5063} {\bibfield  {journal} {\bibinfo  {journal} {Nat. Mater.}\
  }\textbf {\bibinfo {volume} {17}},\ \bibinfo {pages} {119} (\bibinfo {year}
  {2018})}\BibitemShut {NoStop}%
\bibitem [{\citenamefont {Perdigón-Toro}\ \emph {et~al.}(2020)\citenamefont
  {Perdigón-Toro}, \citenamefont {Zhang}, \citenamefont {Markina},
  \citenamefont {Yuan}, \citenamefont {Hosseini}, \citenamefont {Wolff},
  \citenamefont {Zuo}, \citenamefont {Stolterfoht}, \citenamefont {Zou},
  \citenamefont {Gao}, \citenamefont {Andrienko}, \citenamefont {Shoaee},\ and\
  \citenamefont {Neher}}]{Andrienko}%
  \BibitemOpen
  \bibfield  {author} {\bibinfo {author} {\bibfnamefont {L.}~\bibnamefont
  {Perdigón-Toro}}, \bibinfo {author} {\bibfnamefont {H.}~\bibnamefont
  {Zhang}}, \bibinfo {author} {\bibfnamefont {A.}~\bibnamefont {Markina}},
  \bibinfo {author} {\bibfnamefont {J.}~\bibnamefont {Yuan}}, \bibinfo {author}
  {\bibfnamefont {S.~M.}\ \bibnamefont {Hosseini}}, \bibinfo {author}
  {\bibfnamefont {C.~M.}\ \bibnamefont {Wolff}}, \bibinfo {author}
  {\bibfnamefont {G.}~\bibnamefont {Zuo}}, \bibinfo {author} {\bibfnamefont
  {M.}~\bibnamefont {Stolterfoht}}, \bibinfo {author} {\bibfnamefont
  {Y.}~\bibnamefont {Zou}}, \bibinfo {author} {\bibfnamefont {F.}~\bibnamefont
  {Gao}}, \bibinfo {author} {\bibfnamefont {D.}~\bibnamefont {Andrienko}},
  \bibinfo {author} {\bibfnamefont {S.}~\bibnamefont {Shoaee}}, \ and\ \bibinfo
  {author} {\bibfnamefont {D.}~\bibnamefont {Neher}},\ }\href {\doibase
  https://doi.org/10.1002/adma.201906763} {\bibfield  {journal} {\bibinfo
  {journal} {Adv. Mater.}\ }\textbf {\bibinfo {volume} {32}},\ \bibinfo {pages}
  {1906763} (\bibinfo {year} {2020})}\BibitemShut {NoStop}%
\bibitem [{\citenamefont {Coropceanu}\ \emph {et~al.}(2019)\citenamefont
  {Coropceanu}, \citenamefont {Chen}, \citenamefont {Wang}, \citenamefont
  {Zheng},\ and\ \citenamefont {Br{\'e}das}}]{Bredas2019}%
  \BibitemOpen
  \bibfield  {author} {\bibinfo {author} {\bibfnamefont {V.}~\bibnamefont
  {Coropceanu}}, \bibinfo {author} {\bibfnamefont {X.-K.}\ \bibnamefont
  {Chen}}, \bibinfo {author} {\bibfnamefont {T.}~\bibnamefont {Wang}}, \bibinfo
  {author} {\bibfnamefont {Z.}~\bibnamefont {Zheng}}, \ and\ \bibinfo {author}
  {\bibfnamefont {J.-L.}\ \bibnamefont {Br{\'e}das}},\ }\href {\doibase
  10.1038/s41578-019-0137-9} {\bibfield  {journal} {\bibinfo  {journal} {Nat.
  Rev. Mater.}\ }\textbf {\bibinfo {volume} {4}},\ \bibinfo {pages} {689}
  (\bibinfo {year} {2019})}\BibitemShut {NoStop}%
\bibitem [{\citenamefont {Gao}\ \emph {et~al.}(2015)\citenamefont {Gao},
  \citenamefont {Tress}, \citenamefont {Wang},\ and\ \citenamefont
  {Ingan\"as}}]{Gao2015}%
  \BibitemOpen
  \bibfield  {author} {\bibinfo {author} {\bibfnamefont {F.}~\bibnamefont
  {Gao}}, \bibinfo {author} {\bibfnamefont {W.}~\bibnamefont {Tress}}, \bibinfo
  {author} {\bibfnamefont {J.}~\bibnamefont {Wang}}, \ and\ \bibinfo {author}
  {\bibfnamefont {O.}~\bibnamefont {Ingan\"as}},\ }\href {\doibase
  10.1103/PhysRevLett.114.128701} {\bibfield  {journal} {\bibinfo  {journal}
  {Phys. Rev. Lett.}\ }\textbf {\bibinfo {volume} {114}},\ \bibinfo {pages}
  {128701} (\bibinfo {year} {2015})}\BibitemShut {NoStop}%
\bibitem [{\citenamefont {Eisner}\ \emph {et~al.}(2019)\citenamefont {Eisner},
  \citenamefont {Azzouzi}, \citenamefont {Fei}, \citenamefont {Hou},
  \citenamefont {Anthopoulos}, \citenamefont {Dennis}, \citenamefont {Heeney},\
  and\ \citenamefont {Nelson}}]{JennyNelsonJACS}%
  \BibitemOpen
  \bibfield  {author} {\bibinfo {author} {\bibfnamefont {F.~D.}\ \bibnamefont
  {Eisner}}, \bibinfo {author} {\bibfnamefont {M.}~\bibnamefont {Azzouzi}},
  \bibinfo {author} {\bibfnamefont {Z.}~\bibnamefont {Fei}}, \bibinfo {author}
  {\bibfnamefont {X.}~\bibnamefont {Hou}}, \bibinfo {author} {\bibfnamefont
  {T.~D.}\ \bibnamefont {Anthopoulos}}, \bibinfo {author} {\bibfnamefont
  {T.~J.~S.}\ \bibnamefont {Dennis}}, \bibinfo {author} {\bibfnamefont
  {M.}~\bibnamefont {Heeney}}, \ and\ \bibinfo {author} {\bibfnamefont
  {J.}~\bibnamefont {Nelson}},\ }\href {\doibase 10.1021/jacs.9b01465}
  {\bibfield  {journal} {\bibinfo  {journal} {J. Am. Chem. Soc.}\ }\textbf
  {\bibinfo {volume} {141}},\ \bibinfo {pages} {6362} (\bibinfo {year}
  {2019})}\BibitemShut {NoStop}%
\bibitem [{\citenamefont {Classen}\ \emph {et~al.}(2020)\citenamefont
  {Classen}, \citenamefont {Chochos}, \citenamefont {L{\"u}er}, \citenamefont
  {Gregoriou}, \citenamefont {Wortmann}, \citenamefont {Osvet}, \citenamefont
  {Forberich}, \citenamefont {McCulloch}, \citenamefont {Heum{\"u}ller},\ and\
  \citenamefont {Brabec}}]{classen2020role}%
  \BibitemOpen
  \bibfield  {author} {\bibinfo {author} {\bibfnamefont {A.}~\bibnamefont
  {Classen}}, \bibinfo {author} {\bibfnamefont {C.~L.}\ \bibnamefont
  {Chochos}}, \bibinfo {author} {\bibfnamefont {L.}~\bibnamefont {L{\"u}er}},
  \bibinfo {author} {\bibfnamefont {V.~G.}\ \bibnamefont {Gregoriou}}, \bibinfo
  {author} {\bibfnamefont {J.}~\bibnamefont {Wortmann}}, \bibinfo {author}
  {\bibfnamefont {A.}~\bibnamefont {Osvet}}, \bibinfo {author} {\bibfnamefont
  {K.}~\bibnamefont {Forberich}}, \bibinfo {author} {\bibfnamefont
  {I.}~\bibnamefont {McCulloch}}, \bibinfo {author} {\bibfnamefont
  {T.}~\bibnamefont {Heum{\"u}ller}}, \ and\ \bibinfo {author} {\bibfnamefont
  {C.~J.}\ \bibnamefont {Brabec}},\ }\href@noop {} {\bibfield  {journal}
  {\bibinfo  {journal} {Nat. Energy}\ }\textbf {\bibinfo {volume} {5}},\
  \bibinfo {pages} {711} (\bibinfo {year} {2020})}\BibitemShut {NoStop}%
\bibitem [{\citenamefont {Chen}\ \emph {et~al.}(2021)\citenamefont {Chen},
  \citenamefont {Qian}, \citenamefont {Wang}, \citenamefont {Kirchartz},
  \citenamefont {Tress}, \citenamefont {Yao}, \citenamefont {Yuan},
  \citenamefont {H{\"u}lsbeck}, \citenamefont {Zhang}, \citenamefont {Zou},
  \citenamefont {Sun}, \citenamefont {Li}, \citenamefont {Hou}, \citenamefont
  {Ingan{\"a}s}, \citenamefont {Coropceanu}, \citenamefont {Bredas},\ and\
  \citenamefont {Gao}}]{Chen2021}%
  \BibitemOpen
  \bibfield  {author} {\bibinfo {author} {\bibfnamefont {X.-K.}\ \bibnamefont
  {Chen}}, \bibinfo {author} {\bibfnamefont {D.}~\bibnamefont {Qian}}, \bibinfo
  {author} {\bibfnamefont {Y.}~\bibnamefont {Wang}}, \bibinfo {author}
  {\bibfnamefont {T.}~\bibnamefont {Kirchartz}}, \bibinfo {author}
  {\bibfnamefont {W.}~\bibnamefont {Tress}}, \bibinfo {author} {\bibfnamefont
  {H.}~\bibnamefont {Yao}}, \bibinfo {author} {\bibfnamefont {J.}~\bibnamefont
  {Yuan}}, \bibinfo {author} {\bibfnamefont {M.}~\bibnamefont {H{\"u}lsbeck}},
  \bibinfo {author} {\bibfnamefont {M.}~\bibnamefont {Zhang}}, \bibinfo
  {author} {\bibfnamefont {Y.}~\bibnamefont {Zou}}, \bibinfo {author}
  {\bibfnamefont {Y.}~\bibnamefont {Sun}}, \bibinfo {author} {\bibfnamefont
  {Y.}~\bibnamefont {Li}}, \bibinfo {author} {\bibfnamefont {J.}~\bibnamefont
  {Hou}}, \bibinfo {author} {\bibfnamefont {O.}~\bibnamefont {Ingan{\"a}s}},
  \bibinfo {author} {\bibfnamefont {V.}~\bibnamefont {Coropceanu}}, \bibinfo
  {author} {\bibfnamefont {J.-L.}\ \bibnamefont {Bredas}}, \ and\ \bibinfo
  {author} {\bibfnamefont {F.}~\bibnamefont {Gao}},\ }\href {\doibase
  10.1038/s41560-021-00843-4} {\bibfield  {journal} {\bibinfo  {journal} {Nat.
  Energy}\ }\textbf {\bibinfo {volume} {6}},\ \bibinfo {pages} {799} (\bibinfo
  {year} {2021})}\BibitemShut {NoStop}%
\bibitem [{\citenamefont {Chan}\ \emph {et~al.}(2021)\citenamefont {Chan},
  \citenamefont {Ma}, \citenamefont {Zou}, \citenamefont {Xing}, \citenamefont
  {Zhang}, \citenamefont {Yip}, \citenamefont {Taylor}, \citenamefont {He},
  \citenamefont {Wong},\ and\ \citenamefont
  {Chow}}]{Recombination-temperature}%
  \BibitemOpen
  \bibfield  {author} {\bibinfo {author} {\bibfnamefont {C.~C.~S.}\
  \bibnamefont {Chan}}, \bibinfo {author} {\bibfnamefont {C.}~\bibnamefont
  {Ma}}, \bibinfo {author} {\bibfnamefont {X.}~\bibnamefont {Zou}}, \bibinfo
  {author} {\bibfnamefont {Z.}~\bibnamefont {Xing}}, \bibinfo {author}
  {\bibfnamefont {G.}~\bibnamefont {Zhang}}, \bibinfo {author} {\bibfnamefont
  {H.-L.}\ \bibnamefont {Yip}}, \bibinfo {author} {\bibfnamefont {R.~A.}\
  \bibnamefont {Taylor}}, \bibinfo {author} {\bibfnamefont {Y.}~\bibnamefont
  {He}}, \bibinfo {author} {\bibfnamefont {K.~S.}\ \bibnamefont {Wong}}, \ and\
  \bibinfo {author} {\bibfnamefont {P.~C.~Y.}\ \bibnamefont {Chow}},\ }\href
  {\doibase https://doi.org/10.1002/adfm.202107157} {\bibfield  {journal}
  {\bibinfo  {journal} {Adv. Funct. Mater.}\ }\textbf {\bibinfo {volume}
  {31}},\ \bibinfo {pages} {2107157} (\bibinfo {year} {2021})}\BibitemShut
  {NoStop}%
\bibitem [{\citenamefont {Benduhn}\ \emph {et~al.}(2017)\citenamefont
  {Benduhn}, \citenamefont {Tvingstedt}, \citenamefont {Piersimoni},
  \citenamefont {Ullbrich}, \citenamefont {Fan}, \citenamefont {Tropiano},
  \citenamefont {McGarry}, \citenamefont {Zeika}, \citenamefont {Riede},
  \citenamefont {Douglas}, \citenamefont {Barlow}, \citenamefont {Marder},
  \citenamefont {Neher}, \citenamefont {Spoltore},\ and\ \citenamefont
  {Vandewal}}]{Benduhn2017}%
  \BibitemOpen
  \bibfield  {author} {\bibinfo {author} {\bibfnamefont {J.}~\bibnamefont
  {Benduhn}}, \bibinfo {author} {\bibfnamefont {K.}~\bibnamefont {Tvingstedt}},
  \bibinfo {author} {\bibfnamefont {F.}~\bibnamefont {Piersimoni}}, \bibinfo
  {author} {\bibfnamefont {S.}~\bibnamefont {Ullbrich}}, \bibinfo {author}
  {\bibfnamefont {Y.}~\bibnamefont {Fan}}, \bibinfo {author} {\bibfnamefont
  {M.}~\bibnamefont {Tropiano}}, \bibinfo {author} {\bibfnamefont {K.~A.}\
  \bibnamefont {McGarry}}, \bibinfo {author} {\bibfnamefont {O.}~\bibnamefont
  {Zeika}}, \bibinfo {author} {\bibfnamefont {M.~K.}\ \bibnamefont {Riede}},
  \bibinfo {author} {\bibfnamefont {C.~J.}\ \bibnamefont {Douglas}}, \bibinfo
  {author} {\bibfnamefont {S.}~\bibnamefont {Barlow}}, \bibinfo {author}
  {\bibfnamefont {S.~R.}\ \bibnamefont {Marder}}, \bibinfo {author}
  {\bibfnamefont {D.}~\bibnamefont {Neher}}, \bibinfo {author} {\bibfnamefont
  {D.}~\bibnamefont {Spoltore}}, \ and\ \bibinfo {author} {\bibfnamefont
  {K.}~\bibnamefont {Vandewal}},\ }\href {\doibase 10.1038/nenergy.2017.53}
  {\bibfield  {journal} {\bibinfo  {journal} {Nat. Energy}\ }\textbf {\bibinfo
  {volume} {2}},\ \bibinfo {pages} {17053} (\bibinfo {year}
  {2017})}\BibitemShut {NoStop}%
\bibitem [{\citenamefont {Qian}\ \emph {et~al.}(2018)\citenamefont {Qian},
  \citenamefont {Zheng}, \citenamefont {Yao}, \citenamefont {Tress},
  \citenamefont {Hopper}, \citenamefont {Chen}, \citenamefont {Li},
  \citenamefont {Liu}, \citenamefont {Chen}, \citenamefont {Zhang},
  \citenamefont {Liu}, \citenamefont {Gao}, \citenamefont {Ouyang},
  \citenamefont {Jin}, \citenamefont {Pozina}, \citenamefont {Buyanova},
  \citenamefont {Chen}, \citenamefont {Ingan{\"a}s}, \citenamefont
  {Coropceanu}, \citenamefont {Bredas}, \citenamefont {Yan}, \citenamefont
  {Hou}, \citenamefont {Zhang}, \citenamefont {Bakulin},\ and\ \citenamefont
  {Gao}}]{Qian2018}%
  \BibitemOpen
  \bibfield  {author} {\bibinfo {author} {\bibfnamefont {D.}~\bibnamefont
  {Qian}}, \bibinfo {author} {\bibfnamefont {Z.}~\bibnamefont {Zheng}},
  \bibinfo {author} {\bibfnamefont {H.}~\bibnamefont {Yao}}, \bibinfo {author}
  {\bibfnamefont {W.}~\bibnamefont {Tress}}, \bibinfo {author} {\bibfnamefont
  {T.~R.}\ \bibnamefont {Hopper}}, \bibinfo {author} {\bibfnamefont
  {S.}~\bibnamefont {Chen}}, \bibinfo {author} {\bibfnamefont {S.}~\bibnamefont
  {Li}}, \bibinfo {author} {\bibfnamefont {J.}~\bibnamefont {Liu}}, \bibinfo
  {author} {\bibfnamefont {S.}~\bibnamefont {Chen}}, \bibinfo {author}
  {\bibfnamefont {J.}~\bibnamefont {Zhang}}, \bibinfo {author} {\bibfnamefont
  {X.-K.}\ \bibnamefont {Liu}}, \bibinfo {author} {\bibfnamefont
  {B.}~\bibnamefont {Gao}}, \bibinfo {author} {\bibfnamefont {L.}~\bibnamefont
  {Ouyang}}, \bibinfo {author} {\bibfnamefont {Y.}~\bibnamefont {Jin}},
  \bibinfo {author} {\bibfnamefont {G.}~\bibnamefont {Pozina}}, \bibinfo
  {author} {\bibfnamefont {I.~A.}\ \bibnamefont {Buyanova}}, \bibinfo {author}
  {\bibfnamefont {W.~M.}\ \bibnamefont {Chen}}, \bibinfo {author}
  {\bibfnamefont {O.}~\bibnamefont {Ingan{\"a}s}}, \bibinfo {author}
  {\bibfnamefont {V.}~\bibnamefont {Coropceanu}}, \bibinfo {author}
  {\bibfnamefont {J.-L.}\ \bibnamefont {Bredas}}, \bibinfo {author}
  {\bibfnamefont {H.}~\bibnamefont {Yan}}, \bibinfo {author} {\bibfnamefont
  {J.}~\bibnamefont {Hou}}, \bibinfo {author} {\bibfnamefont {F.}~\bibnamefont
  {Zhang}}, \bibinfo {author} {\bibfnamefont {A.~A.}\ \bibnamefont {Bakulin}},
  \ and\ \bibinfo {author} {\bibfnamefont {F.}~\bibnamefont {Gao}},\ }\href
  {\doibase 10.1038/s41563-018-0128-z} {\bibfield  {journal} {\bibinfo
  {journal} {Nat. Mater.}\ }\textbf {\bibinfo {volume} {17}},\ \bibinfo {pages}
  {703} (\bibinfo {year} {2018})}\BibitemShut {NoStop}%
\bibitem [{\citenamefont {Hofinger}\ \emph {et~al.}(2021)\citenamefont
  {Hofinger}, \citenamefont {Putz}, \citenamefont {Mayr}, \citenamefont
  {Gugujonovic}, \citenamefont {Wielend},\ and\ \citenamefont
  {Scharber}}]{Hofinger}%
  \BibitemOpen
  \bibfield  {author} {\bibinfo {author} {\bibfnamefont {J.}~\bibnamefont
  {Hofinger}}, \bibinfo {author} {\bibfnamefont {C.}~\bibnamefont {Putz}},
  \bibinfo {author} {\bibfnamefont {F.}~\bibnamefont {Mayr}}, \bibinfo {author}
  {\bibfnamefont {K.}~\bibnamefont {Gugujonovic}}, \bibinfo {author}
  {\bibfnamefont {D.}~\bibnamefont {Wielend}}, \ and\ \bibinfo {author}
  {\bibfnamefont {M.~C.}\ \bibnamefont {Scharber}},\ }\href {\doibase
  10.1039/D1MA00293G} {\bibfield  {journal} {\bibinfo  {journal} {Mater. Adv.}\
  }\textbf {\bibinfo {volume} {2}},\ \bibinfo {pages} {4291} (\bibinfo {year}
  {2021})}\BibitemShut {NoStop}%
\bibitem [{\citenamefont {Chen}\ \emph {et~al.}(2018)\citenamefont {Chen},
  \citenamefont {Coropceanu},\ and\ \citenamefont {Br{\'e}das}}]{Chen2018}%
  \BibitemOpen
  \bibfield  {author} {\bibinfo {author} {\bibfnamefont {X.-K.}\ \bibnamefont
  {Chen}}, \bibinfo {author} {\bibfnamefont {V.}~\bibnamefont {Coropceanu}}, \
  and\ \bibinfo {author} {\bibfnamefont {J.-L.}\ \bibnamefont {Br{\'e}das}},\
  }\href {\doibase 10.1038/s41467-018-07707-8} {\bibfield  {journal} {\bibinfo
  {journal} {Nat. Commun.}\ }\textbf {\bibinfo {volume} {9}},\ \bibinfo {pages}
  {5295} (\bibinfo {year} {2018})}\BibitemShut {NoStop}%
\bibitem [{\citenamefont {Coffey}\ \emph {et~al.}(2012)\citenamefont {Coffey},
  \citenamefont {Larson}, \citenamefont {Hains}, \citenamefont {Whitaker},
  \citenamefont {Kopidakis}, \citenamefont {Boltalina}, \citenamefont
  {Strauss},\ and\ \citenamefont {Rumbles}}]{Coffey}%
  \BibitemOpen
  \bibfield  {author} {\bibinfo {author} {\bibfnamefont {D.~C.}\ \bibnamefont
  {Coffey}}, \bibinfo {author} {\bibfnamefont {B.~W.}\ \bibnamefont {Larson}},
  \bibinfo {author} {\bibfnamefont {A.~W.}\ \bibnamefont {Hains}}, \bibinfo
  {author} {\bibfnamefont {J.~B.}\ \bibnamefont {Whitaker}}, \bibinfo {author}
  {\bibfnamefont {N.}~\bibnamefont {Kopidakis}}, \bibinfo {author}
  {\bibfnamefont {O.~V.}\ \bibnamefont {Boltalina}}, \bibinfo {author}
  {\bibfnamefont {S.~H.}\ \bibnamefont {Strauss}}, \ and\ \bibinfo {author}
  {\bibfnamefont {G.}~\bibnamefont {Rumbles}},\ }\href {\doibase
  10.1021/jp302275z} {\bibfield  {journal} {\bibinfo  {journal} {J. Phys. Chem.
  C}\ }\textbf {\bibinfo {volume} {116}},\ \bibinfo {pages} {8916} (\bibinfo
  {year} {2012})}\BibitemShut {NoStop}%
\bibitem [{\citenamefont {Ward}\ \emph {et~al.}(2015)\citenamefont {Ward},
  \citenamefont {Ruseckas}, \citenamefont {Kareem}, \citenamefont {Ebenhoch},
  \citenamefont {Serrano}, \citenamefont {Al-Eid}, \citenamefont {Fitzpatrick},
  \citenamefont {Rotello}, \citenamefont {Cooke},\ and\ \citenamefont
  {Samuel}}]{Ward}%
  \BibitemOpen
  \bibfield  {author} {\bibinfo {author} {\bibfnamefont {A.~J.}\ \bibnamefont
  {Ward}}, \bibinfo {author} {\bibfnamefont {A.}~\bibnamefont {Ruseckas}},
  \bibinfo {author} {\bibfnamefont {M.~M.}\ \bibnamefont {Kareem}}, \bibinfo
  {author} {\bibfnamefont {B.}~\bibnamefont {Ebenhoch}}, \bibinfo {author}
  {\bibfnamefont {L.~A.}\ \bibnamefont {Serrano}}, \bibinfo {author}
  {\bibfnamefont {M.}~\bibnamefont {Al-Eid}}, \bibinfo {author} {\bibfnamefont
  {B.}~\bibnamefont {Fitzpatrick}}, \bibinfo {author} {\bibfnamefont {V.~M.}\
  \bibnamefont {Rotello}}, \bibinfo {author} {\bibfnamefont {G.}~\bibnamefont
  {Cooke}}, \ and\ \bibinfo {author} {\bibfnamefont {I.~D.~W.}\ \bibnamefont
  {Samuel}},\ }\href {\doibase https://doi.org/10.1002/adma.201405623}
  {\bibfield  {journal} {\bibinfo  {journal} {Adv. Mat.}\ }\textbf {\bibinfo
  {volume} {27}},\ \bibinfo {pages} {2496} (\bibinfo {year}
  {2015})}\BibitemShut {NoStop}%
\bibitem [{\citenamefont {Rand}\ \emph {et~al.}(2007)\citenamefont {Rand},
  \citenamefont {Burk},\ and\ \citenamefont {Forrest}}]{Forrest}%
  \BibitemOpen
  \bibfield  {author} {\bibinfo {author} {\bibfnamefont {B.~P.}\ \bibnamefont
  {Rand}}, \bibinfo {author} {\bibfnamefont {D.~P.}\ \bibnamefont {Burk}}, \
  and\ \bibinfo {author} {\bibfnamefont {S.~R.}\ \bibnamefont {Forrest}},\
  }\href {\doibase 10.1103/PhysRevB.75.115327} {\bibfield  {journal} {\bibinfo
  {journal} {Phys. Rev. B}\ }\textbf {\bibinfo {volume} {75}},\ \bibinfo
  {pages} {115327} (\bibinfo {year} {2007})}\BibitemShut {NoStop}%
\bibitem [{\citenamefont {Redfield}(1957)}]{redfield1957theory}%
  \BibitemOpen
  \bibfield  {author} {\bibinfo {author} {\bibfnamefont {A.~G.}\ \bibnamefont
  {Redfield}},\ }\href@noop {} {\bibfield  {journal} {\bibinfo  {journal} {IBM
  J. Res. Dev.}\ }\textbf {\bibinfo {volume} {1}},\ \bibinfo {pages} {19}
  (\bibinfo {year} {1957})}\BibitemShut {NoStop}%
\bibitem [{\citenamefont {Redfield}(1965)}]{redfield1965theory}%
  \BibitemOpen
  \bibfield  {author} {\bibinfo {author} {\bibfnamefont {A.}~\bibnamefont
  {Redfield}},\ }in\ \href@noop {} {\emph {\bibinfo {booktitle} {Advances in
  Magnetic and Optical Resonance}}},\ Vol.~\bibinfo {volume} {1}\ (\bibinfo
  {publisher} {Elsevier},\ \bibinfo {year} {1965})\ pp.\ \bibinfo {pages}
  {1--32}\BibitemShut {NoStop}%
\bibitem [{\citenamefont {Andermann}\ and\ \citenamefont {Rego}(2022)}]{Artur}%
  \BibitemOpen
  \bibfield  {author} {\bibinfo {author} {\bibfnamefont {A.~M.}\ \bibnamefont
  {Andermann}}\ and\ \bibinfo {author} {\bibfnamefont {L.~G.~C.}\ \bibnamefont
  {Rego}},\ }\href {\doibase 10.1063/5.0076611} {\bibfield  {journal} {\bibinfo
   {journal} {J. Chem. Phys.}\ }\textbf {\bibinfo {volume} {156}},\ \bibinfo
  {pages} {024104} (\bibinfo {year} {2022})}\BibitemShut {NoStop}%
\bibitem [{\citenamefont {Candiotto}\ \emph {et~al.}(2017)\citenamefont
  {Candiotto}, \citenamefont {Torres}, \citenamefont {Mazon},\ and\
  \citenamefont {Rego}}]{candiotto2017charge}%
  \BibitemOpen
  \bibfield  {author} {\bibinfo {author} {\bibfnamefont {G.}~\bibnamefont
  {Candiotto}}, \bibinfo {author} {\bibfnamefont {A.}~\bibnamefont {Torres}},
  \bibinfo {author} {\bibfnamefont {K.~T.}\ \bibnamefont {Mazon}}, \ and\
  \bibinfo {author} {\bibfnamefont {L.~G.~C.}\ \bibnamefont {Rego}},\
  }\href@noop {} {\bibfield  {journal} {\bibinfo  {journal} {J. Phys. Chem. C}\
  }\textbf {\bibinfo {volume} {121}},\ \bibinfo {pages} {23276} (\bibinfo
  {year} {2017})}\BibitemShut {NoStop}%
\bibitem [{\citenamefont {Credgington}\ \emph {et~al.}(2012)\citenamefont
  {Credgington}, \citenamefont {Jamieson}, \citenamefont {Walker},
  \citenamefont {Nguyen},\ and\ \citenamefont {Durrant}}]{Durrant2012}%
  \BibitemOpen
  \bibfield  {author} {\bibinfo {author} {\bibfnamefont {D.}~\bibnamefont
  {Credgington}}, \bibinfo {author} {\bibfnamefont {F.~C.}\ \bibnamefont
  {Jamieson}}, \bibinfo {author} {\bibfnamefont {B.}~\bibnamefont {Walker}},
  \bibinfo {author} {\bibfnamefont {T.-Q.}\ \bibnamefont {Nguyen}}, \ and\
  \bibinfo {author} {\bibfnamefont {J.~R.}\ \bibnamefont {Durrant}},\ }\href
  {\doibase https://doi.org/10.1002/adma.201104738} {\bibfield  {journal}
  {\bibinfo  {journal} {Adv. Mater.}\ }\textbf {\bibinfo {volume} {24}},\
  \bibinfo {pages} {2135} (\bibinfo {year} {2012})}\BibitemShut {NoStop}%
\bibitem [{\citenamefont {Haberkorn}(1976)}]{Haberkorn}%
  \BibitemOpen
  \bibfield  {author} {\bibinfo {author} {\bibfnamefont {R.}~\bibnamefont
  {Haberkorn}},\ }\href {\doibase 10.1080/00268977600102851} {\bibfield
  {journal} {\bibinfo  {journal} {Mol. Phys.}\ }\textbf {\bibinfo {volume}
  {32}},\ \bibinfo {pages} {1491} (\bibinfo {year} {1976})}\BibitemShut
  {NoStop}%
\bibitem [{\citenamefont {Rebentrost}\ \emph {et~al.}(2009)\citenamefont
  {Rebentrost}, \citenamefont {Mohseni},\ and\ \citenamefont
  {Aspuru-Guzik}}]{Aspuru}%
  \BibitemOpen
  \bibfield  {author} {\bibinfo {author} {\bibfnamefont {P.}~\bibnamefont
  {Rebentrost}}, \bibinfo {author} {\bibfnamefont {M.}~\bibnamefont {Mohseni}},
  \ and\ \bibinfo {author} {\bibfnamefont {A.}~\bibnamefont {Aspuru-Guzik}},\
  }\href {\doibase 10.1021/jp901724d} {\bibfield  {journal} {\bibinfo
  {journal} {The Journal of Physical Chemistry B}\ }\textbf {\bibinfo {volume}
  {113}},\ \bibinfo {pages} {9942} (\bibinfo {year} {2009})},\ \bibinfo {note}
  {pMID: 19603843}\BibitemShut {NoStop}%
\bibitem [{\citenamefont {Burke}\ \emph {et~al.}(2015)\citenamefont {Burke},
  \citenamefont {Sweetnam}, \citenamefont {Vandewal},\ and\ \citenamefont
  {McGehee}}]{Burke}%
  \BibitemOpen
  \bibfield  {author} {\bibinfo {author} {\bibfnamefont {T.~M.}\ \bibnamefont
  {Burke}}, \bibinfo {author} {\bibfnamefont {S.}~\bibnamefont {Sweetnam}},
  \bibinfo {author} {\bibfnamefont {K.}~\bibnamefont {Vandewal}}, \ and\
  \bibinfo {author} {\bibfnamefont {M.~D.}\ \bibnamefont {McGehee}},\ }\href
  {\doibase https://doi.org/10.1002/aenm.201500123} {\bibfield  {journal}
  {\bibinfo  {journal} {Adv. Energy Mater.}\ }\textbf {\bibinfo {volume} {5}},\
  \bibinfo {pages} {1500123} (\bibinfo {year} {2015})}\BibitemShut {NoStop}%
\bibitem [{\citenamefont {Lee}\ \emph {et~al.}(2015)\citenamefont {Lee},
  \citenamefont {Arag{\'o}},\ and\ \citenamefont {Troisi}}]{lee2015charge}%
  \BibitemOpen
  \bibfield  {author} {\bibinfo {author} {\bibfnamefont {M.~H.}\ \bibnamefont
  {Lee}}, \bibinfo {author} {\bibfnamefont {J.}~\bibnamefont {Arag{\'o}}}, \
  and\ \bibinfo {author} {\bibfnamefont {A.}~\bibnamefont {Troisi}},\
  }\href@noop {} {\bibfield  {journal} {\bibinfo  {journal} {J. Phys. Chem. C}\
  }\textbf {\bibinfo {volume} {119}},\ \bibinfo {pages} {14989} (\bibinfo
  {year} {2015})}\BibitemShut {NoStop}%
\bibitem [{\citenamefont {Coropceanu}\ \emph {et~al.}(2007)\citenamefont
  {Coropceanu}, \citenamefont {Cornil}, \citenamefont {da~Silva~Filho},
  \citenamefont {Olivier}, \citenamefont {Silbey},\ and\ \citenamefont
  {Br{\'e}das}}]{coropceanu2007charge}%
  \BibitemOpen
  \bibfield  {author} {\bibinfo {author} {\bibfnamefont {V.}~\bibnamefont
  {Coropceanu}}, \bibinfo {author} {\bibfnamefont {J.}~\bibnamefont {Cornil}},
  \bibinfo {author} {\bibfnamefont {D.~A.}\ \bibnamefont {da~Silva~Filho}},
  \bibinfo {author} {\bibfnamefont {Y.}~\bibnamefont {Olivier}}, \bibinfo
  {author} {\bibfnamefont {R.}~\bibnamefont {Silbey}}, \ and\ \bibinfo {author}
  {\bibfnamefont {J.-L.}\ \bibnamefont {Br{\'e}das}},\ }\href@noop {}
  {\bibfield  {journal} {\bibinfo  {journal} {Chem. Rev.}\ }\textbf {\bibinfo
  {volume} {107}},\ \bibinfo {pages} {926} (\bibinfo {year}
  {2007})}\BibitemShut {NoStop}%
\bibitem [{\citenamefont {Mozafari}\ and\ \citenamefont
  {Stafstr{\"o}m}(2013)}]{mozafari2013polaron}%
  \BibitemOpen
  \bibfield  {author} {\bibinfo {author} {\bibfnamefont {E.}~\bibnamefont
  {Mozafari}}\ and\ \bibinfo {author} {\bibfnamefont {S.}~\bibnamefont
  {Stafstr{\"o}m}},\ }\href@noop {} {\bibfield  {journal} {\bibinfo  {journal}
  {J. Chem. Phys.}\ }\textbf {\bibinfo {volume} {138}},\ \bibinfo {pages}
  {184104} (\bibinfo {year} {2013})}\BibitemShut {NoStop}%
\bibitem [{\citenamefont {Kocherzhenko}\ \emph {et~al.}(2015)\citenamefont
  {Kocherzhenko}, \citenamefont {Lee}, \citenamefont {Forsuelo},\ and\
  \citenamefont {Whaley}}]{kocherzhenko2015coherent}%
  \BibitemOpen
  \bibfield  {author} {\bibinfo {author} {\bibfnamefont {A.~A.}\ \bibnamefont
  {Kocherzhenko}}, \bibinfo {author} {\bibfnamefont {D.}~\bibnamefont {Lee}},
  \bibinfo {author} {\bibfnamefont {M.~A.}\ \bibnamefont {Forsuelo}}, \ and\
  \bibinfo {author} {\bibfnamefont {K.~B.}\ \bibnamefont {Whaley}},\
  }\href@noop {} {\bibfield  {journal} {\bibinfo  {journal} {J. Phys. Chem. C}\
  }\textbf {\bibinfo {volume} {119}},\ \bibinfo {pages} {7590} (\bibinfo {year}
  {2015})}\BibitemShut {NoStop}%
\bibitem [{\citenamefont {De~Sio}\ and\ \citenamefont
  {Lienau}(2017)}]{de2017vibronic}%
  \BibitemOpen
  \bibfield  {author} {\bibinfo {author} {\bibfnamefont {A.}~\bibnamefont
  {De~Sio}}\ and\ \bibinfo {author} {\bibfnamefont {C.}~\bibnamefont
  {Lienau}},\ }\href@noop {} {\bibfield  {journal} {\bibinfo  {journal} {Phys.
  Chem. Chem. Phys.}\ }\textbf {\bibinfo {volume} {19}},\ \bibinfo {pages}
  {18813} (\bibinfo {year} {2017})}\BibitemShut {NoStop}%
\bibitem [{\citenamefont {Troisi}\ and\ \citenamefont
  {Orlandi}(2006)}]{troisi2006dynamics}%
  \BibitemOpen
  \bibfield  {author} {\bibinfo {author} {\bibfnamefont {A.}~\bibnamefont
  {Troisi}}\ and\ \bibinfo {author} {\bibfnamefont {G.}~\bibnamefont
  {Orlandi}},\ }\href@noop {} {\bibfield  {journal} {\bibinfo  {journal} {J.
  Phys. Chem. A}\ }\textbf {\bibinfo {volume} {110}},\ \bibinfo {pages} {4065}
  (\bibinfo {year} {2006})}\BibitemShut {NoStop}%
\bibitem [{\citenamefont {Vandewal}\ \emph {et~al.}(2017)\citenamefont
  {Vandewal}, \citenamefont {Benduhn}, \citenamefont {Schellhammer},
  \citenamefont {Vangerven}, \citenamefont {Rückert}, \citenamefont
  {Piersimoni}, \citenamefont {Scholz}, \citenamefont {Zeika}, \citenamefont
  {Fan}, \citenamefont {Barlow}, \citenamefont {Neher}, \citenamefont {Marder},
  \citenamefont {Manca}, \citenamefont {Spoltore}, \citenamefont {Cuniberti},\
  and\ \citenamefont {Ortmann}}]{Vandewal2017}%
  \BibitemOpen
  \bibfield  {author} {\bibinfo {author} {\bibfnamefont {K.}~\bibnamefont
  {Vandewal}}, \bibinfo {author} {\bibfnamefont {J.}~\bibnamefont {Benduhn}},
  \bibinfo {author} {\bibfnamefont {K.~S.}\ \bibnamefont {Schellhammer}},
  \bibinfo {author} {\bibfnamefont {T.}~\bibnamefont {Vangerven}}, \bibinfo
  {author} {\bibfnamefont {J.~E.}\ \bibnamefont {Rückert}}, \bibinfo {author}
  {\bibfnamefont {F.}~\bibnamefont {Piersimoni}}, \bibinfo {author}
  {\bibfnamefont {R.}~\bibnamefont {Scholz}}, \bibinfo {author} {\bibfnamefont
  {O.}~\bibnamefont {Zeika}}, \bibinfo {author} {\bibfnamefont
  {Y.}~\bibnamefont {Fan}}, \bibinfo {author} {\bibfnamefont {S.}~\bibnamefont
  {Barlow}}, \bibinfo {author} {\bibfnamefont {D.}~\bibnamefont {Neher}},
  \bibinfo {author} {\bibfnamefont {S.~R.}\ \bibnamefont {Marder}}, \bibinfo
  {author} {\bibfnamefont {J.}~\bibnamefont {Manca}}, \bibinfo {author}
  {\bibfnamefont {D.}~\bibnamefont {Spoltore}}, \bibinfo {author}
  {\bibfnamefont {G.}~\bibnamefont {Cuniberti}}, \ and\ \bibinfo {author}
  {\bibfnamefont {F.}~\bibnamefont {Ortmann}},\ }\href {\doibase
  10.1021/jacs.6b12857} {\bibfield  {journal} {\bibinfo  {journal} {J. Am.
  Chem. Soc.}\ }\textbf {\bibinfo {volume} {139}},\ \bibinfo {pages} {1699}
  (\bibinfo {year} {2017})}\BibitemShut {NoStop}%
\bibitem [{\citenamefont {Mikhnenko}\ \emph {et~al.}(2012)\citenamefont
  {Mikhnenko}, \citenamefont {Azimi}, \citenamefont {Scharber}, \citenamefont
  {Morana}, \citenamefont {Blom},\ and\ \citenamefont {Loi}}]{Exc-diffusion}%
  \BibitemOpen
  \bibfield  {author} {\bibinfo {author} {\bibfnamefont {O.~V.}\ \bibnamefont
  {Mikhnenko}}, \bibinfo {author} {\bibfnamefont {H.}~\bibnamefont {Azimi}},
  \bibinfo {author} {\bibfnamefont {M.}~\bibnamefont {Scharber}}, \bibinfo
  {author} {\bibfnamefont {M.}~\bibnamefont {Morana}}, \bibinfo {author}
  {\bibfnamefont {P.~W.~M.}\ \bibnamefont {Blom}}, \ and\ \bibinfo {author}
  {\bibfnamefont {M.~A.}\ \bibnamefont {Loi}},\ }\href {\doibase
  10.1039/C2EE03466B} {\bibfield  {journal} {\bibinfo  {journal} {Energy
  Environ. Sci.}\ }\textbf {\bibinfo {volume} {5}},\ \bibinfo {pages} {6960}
  (\bibinfo {year} {2012})}\BibitemShut {NoStop}%
\bibitem [{\citenamefont {Firdaus}\ \emph {et~al.}(2020)\citenamefont
  {Firdaus}, \citenamefont {Le~Corre}, \citenamefont {Karuthedath},
  \citenamefont {Liu}, \citenamefont {Markina}, \citenamefont {Huang},
  \citenamefont {Chattopadhyay}, \citenamefont {Nahid}, \citenamefont
  {Nugraha}, \citenamefont {Lin}, \citenamefont {Seitkhan}, \citenamefont
  {Basu}, \citenamefont {Zhang}, \citenamefont {McCulloch}, \citenamefont
  {Ade}, \citenamefont {Labram}, \citenamefont {Laquai}, \citenamefont
  {Andrienko}, \citenamefont {Koster},\ and\ \citenamefont
  {Anthopoulos}}]{Firdaus2020}%
  \BibitemOpen
  \bibfield  {author} {\bibinfo {author} {\bibfnamefont {Y.}~\bibnamefont
  {Firdaus}}, \bibinfo {author} {\bibfnamefont {V.~M.}\ \bibnamefont
  {Le~Corre}}, \bibinfo {author} {\bibfnamefont {S.}~\bibnamefont
  {Karuthedath}}, \bibinfo {author} {\bibfnamefont {W.}~\bibnamefont {Liu}},
  \bibinfo {author} {\bibfnamefont {A.}~\bibnamefont {Markina}}, \bibinfo
  {author} {\bibfnamefont {W.}~\bibnamefont {Huang}}, \bibinfo {author}
  {\bibfnamefont {S.}~\bibnamefont {Chattopadhyay}}, \bibinfo {author}
  {\bibfnamefont {M.~M.}\ \bibnamefont {Nahid}}, \bibinfo {author}
  {\bibfnamefont {M.~I.}\ \bibnamefont {Nugraha}}, \bibinfo {author}
  {\bibfnamefont {Y.}~\bibnamefont {Lin}}, \bibinfo {author} {\bibfnamefont
  {A.}~\bibnamefont {Seitkhan}}, \bibinfo {author} {\bibfnamefont
  {A.}~\bibnamefont {Basu}}, \bibinfo {author} {\bibfnamefont {W.}~\bibnamefont
  {Zhang}}, \bibinfo {author} {\bibfnamefont {I.}~\bibnamefont {McCulloch}},
  \bibinfo {author} {\bibfnamefont {H.}~\bibnamefont {Ade}}, \bibinfo {author}
  {\bibfnamefont {J.}~\bibnamefont {Labram}}, \bibinfo {author} {\bibfnamefont
  {F.}~\bibnamefont {Laquai}}, \bibinfo {author} {\bibfnamefont
  {D.}~\bibnamefont {Andrienko}}, \bibinfo {author} {\bibfnamefont {L.~J.~A.}\
  \bibnamefont {Koster}}, \ and\ \bibinfo {author} {\bibfnamefont {T.~D.}\
  \bibnamefont {Anthopoulos}},\ }\href {\doibase 10.1038/s41467-020-19029-9}
  {\bibfield  {journal} {\bibinfo  {journal} {Nat. Commun.}\ }\textbf {\bibinfo
  {volume} {11}},\ \bibinfo {pages} {5220} (\bibinfo {year}
  {2020})}\BibitemShut {NoStop}%
\bibitem [{\citenamefont {Dong}\ \emph {et~al.}(2019)\citenamefont {Dong},
  \citenamefont {Cha}, \citenamefont {Zhang}, \citenamefont {Pastor},
  \citenamefont {Tuladhar}, \citenamefont {McCulloch}, \citenamefont
  {Durrant},\ and\ \citenamefont {Bakulin}}]{Dong2019}%
  \BibitemOpen
  \bibfield  {author} {\bibinfo {author} {\bibfnamefont {Y.}~\bibnamefont
  {Dong}}, \bibinfo {author} {\bibfnamefont {H.}~\bibnamefont {Cha}}, \bibinfo
  {author} {\bibfnamefont {J.}~\bibnamefont {Zhang}}, \bibinfo {author}
  {\bibfnamefont {E.}~\bibnamefont {Pastor}}, \bibinfo {author} {\bibfnamefont
  {P.~S.}\ \bibnamefont {Tuladhar}}, \bibinfo {author} {\bibfnamefont
  {I.}~\bibnamefont {McCulloch}}, \bibinfo {author} {\bibfnamefont {J.~R.}\
  \bibnamefont {Durrant}}, \ and\ \bibinfo {author} {\bibfnamefont {A.~A.}\
  \bibnamefont {Bakulin}},\ }\href {\doibase 10.1063/1.5079285} {\bibfield
  {journal} {\bibinfo  {journal} {J. Chem. Phys.}\ }\textbf {\bibinfo {volume}
  {150}},\ \bibinfo {pages} {104704} (\bibinfo {year} {2019})}\BibitemShut
  {NoStop}%
\bibitem [{\citenamefont {Unger}\ \emph {et~al.}(2017)\citenamefont {Unger},
  \citenamefont {Wedler}, \citenamefont {Kahle}, \citenamefont {Scherf},
  \citenamefont {Bässler},\ and\ \citenamefont {Köhler}}]{Unger}%
  \BibitemOpen
  \bibfield  {author} {\bibinfo {author} {\bibfnamefont {T.}~\bibnamefont
  {Unger}}, \bibinfo {author} {\bibfnamefont {S.}~\bibnamefont {Wedler}},
  \bibinfo {author} {\bibfnamefont {F.-J.}\ \bibnamefont {Kahle}}, \bibinfo
  {author} {\bibfnamefont {U.}~\bibnamefont {Scherf}}, \bibinfo {author}
  {\bibfnamefont {H.}~\bibnamefont {Bässler}}, \ and\ \bibinfo {author}
  {\bibfnamefont {A.}~\bibnamefont {Köhler}},\ }\href {\doibase
  10.1021/acs.jpcc.7b09213} {\bibfield  {journal} {\bibinfo  {journal} {J.
  Phys. Chem. C}\ }\textbf {\bibinfo {volume} {121}},\ \bibinfo {pages} {22739}
  (\bibinfo {year} {2017})}\BibitemShut {NoStop}%
\bibitem [{\citenamefont {Jortner}(1976)}]{Jortner}%
  \BibitemOpen
  \bibfield  {author} {\bibinfo {author} {\bibfnamefont {J.}~\bibnamefont
  {Jortner}},\ }\href {\doibase 10.1063/1.432142} {\bibfield  {journal}
  {\bibinfo  {journal} {J. Chem. Phys.}\ }\textbf {\bibinfo {volume} {64}},\
  \bibinfo {pages} {4860} (\bibinfo {year} {1976})}\BibitemShut {NoStop}%
\bibitem [{\citenamefont {Barbara}\ \emph {et~al.}(1996)\citenamefont
  {Barbara}, \citenamefont {Meyer},\ and\ \citenamefont {Ratner}}]{MJL}%
  \BibitemOpen
  \bibfield  {author} {\bibinfo {author} {\bibfnamefont {P.~F.}\ \bibnamefont
  {Barbara}}, \bibinfo {author} {\bibfnamefont {T.~J.}\ \bibnamefont {Meyer}},
  \ and\ \bibinfo {author} {\bibfnamefont {M.~A.}\ \bibnamefont {Ratner}},\
  }\href {\doibase 10.1021/jp9605663} {\bibfield  {journal} {\bibinfo
  {journal} {J. Phys. Chem.}\ }\textbf {\bibinfo {volume} {100}},\ \bibinfo
  {pages} {13148} (\bibinfo {year} {1996})}\BibitemShut {NoStop}%
\bibitem [{\citenamefont {Chaudhuri}\ \emph {et~al.}(2017)\citenamefont
  {Chaudhuri}, \citenamefont {Hedström}, \citenamefont {Méndez-Hernández},
  \citenamefont {Hendrickson}, \citenamefont {Jung}, \citenamefont {Ho},\ and\
  \citenamefont {Batista}}]{VSBatista}%
  \BibitemOpen
  \bibfield  {author} {\bibinfo {author} {\bibfnamefont {S.}~\bibnamefont
  {Chaudhuri}}, \bibinfo {author} {\bibfnamefont {S.}~\bibnamefont
  {Hedström}}, \bibinfo {author} {\bibfnamefont {D.~D.}\ \bibnamefont
  {Méndez-Hernández}}, \bibinfo {author} {\bibfnamefont {H.~P.}\ \bibnamefont
  {Hendrickson}}, \bibinfo {author} {\bibfnamefont {K.~A.}\ \bibnamefont
  {Jung}}, \bibinfo {author} {\bibfnamefont {J.}~\bibnamefont {Ho}}, \ and\
  \bibinfo {author} {\bibfnamefont {V.~S.}\ \bibnamefont {Batista}},\ }\href
  {\doibase 10.1021/acs.jctc.7b00513} {\bibfield  {journal} {\bibinfo
  {journal} {J. Chem. Theory Comput.}\ }\textbf {\bibinfo {volume} {13}},\
  \bibinfo {pages} {6000} (\bibinfo {year} {2017})}\BibitemShut {NoStop}%
\bibitem [{\citenamefont {Proctor}\ and\ \citenamefont
  {Nguyen}(2015)}]{Proctor-leakage}%
  \BibitemOpen
  \bibfield  {author} {\bibinfo {author} {\bibfnamefont {C.~M.}\ \bibnamefont
  {Proctor}}\ and\ \bibinfo {author} {\bibfnamefont {T.-Q.}\ \bibnamefont
  {Nguyen}},\ }\href {\doibase 10.1063/1.4913589} {\bibfield  {journal}
  {\bibinfo  {journal} {Appl. Phys. Lett.}\ }\textbf {\bibinfo {volume}
  {106}},\ \bibinfo {pages} {083301} (\bibinfo {year} {2015})}\BibitemShut
  {NoStop}%
\bibitem [{\citenamefont {Rauh}\ \emph {et~al.}(2011)\citenamefont {Rauh},
  \citenamefont {Wagenpfahl}, \citenamefont {Deibel},\ and\ \citenamefont
  {Dyakonov}}]{Rauh}%
  \BibitemOpen
  \bibfield  {author} {\bibinfo {author} {\bibfnamefont {D.}~\bibnamefont
  {Rauh}}, \bibinfo {author} {\bibfnamefont {A.}~\bibnamefont {Wagenpfahl}},
  \bibinfo {author} {\bibfnamefont {C.}~\bibnamefont {Deibel}}, \ and\ \bibinfo
  {author} {\bibfnamefont {V.}~\bibnamefont {Dyakonov}},\ }\href {\doibase
  10.1063/1.3566979} {\bibfield  {journal} {\bibinfo  {journal} {Appl. Phys.
  Lett.}\ }\textbf {\bibinfo {volume} {98}},\ \bibinfo {pages} {133301}
  (\bibinfo {year} {2011})}\BibitemShut {NoStop}%
\bibitem [{\citenamefont {Vandewal}\ \emph {et~al.}(2010)\citenamefont
  {Vandewal}, \citenamefont {Tvingstedt}, \citenamefont {Gadisa}, \citenamefont
  {Ingan\"as},\ and\ \citenamefont {Manca}}]{Inganas-2010}%
  \BibitemOpen
  \bibfield  {author} {\bibinfo {author} {\bibfnamefont {K.}~\bibnamefont
  {Vandewal}}, \bibinfo {author} {\bibfnamefont {K.}~\bibnamefont
  {Tvingstedt}}, \bibinfo {author} {\bibfnamefont {A.}~\bibnamefont {Gadisa}},
  \bibinfo {author} {\bibfnamefont {O.}~\bibnamefont {Ingan\"as}}, \ and\
  \bibinfo {author} {\bibfnamefont {J.~V.}\ \bibnamefont {Manca}},\ }\href
  {\doibase 10.1103/PhysRevB.81.125204} {\bibfield  {journal} {\bibinfo
  {journal} {Phys. Rev. B}\ }\textbf {\bibinfo {volume} {81}},\ \bibinfo
  {pages} {125204} (\bibinfo {year} {2010})}\BibitemShut {NoStop}%
\bibitem [{\citenamefont {Proctor}\ \emph {et~al.}(2014)\citenamefont
  {Proctor}, \citenamefont {Albrecht}, \citenamefont {Kuik}, \citenamefont
  {Neher},\ and\ \citenamefont {Nguyen}}]{Proctor}%
  \BibitemOpen
  \bibfield  {author} {\bibinfo {author} {\bibfnamefont {C.~M.}\ \bibnamefont
  {Proctor}}, \bibinfo {author} {\bibfnamefont {S.}~\bibnamefont {Albrecht}},
  \bibinfo {author} {\bibfnamefont {M.}~\bibnamefont {Kuik}}, \bibinfo {author}
  {\bibfnamefont {D.}~\bibnamefont {Neher}}, \ and\ \bibinfo {author}
  {\bibfnamefont {T.-Q.}\ \bibnamefont {Nguyen}},\ }\href {\doibase
  https://doi.org/10.1002/aenm.201400230} {\bibfield  {journal} {\bibinfo
  {journal} {Adv. Energy Mater.}\ }\textbf {\bibinfo {volume} {4}},\ \bibinfo
  {pages} {1400230} (\bibinfo {year} {2014})}\BibitemShut {NoStop}%
\bibitem [{\citenamefont {Egorova}\ \emph {et~al.}(2001)\citenamefont
  {Egorova}, \citenamefont {K{\"u}hl},\ and\ \citenamefont
  {Domcke}}]{egorova2001modeling}%
  \BibitemOpen
  \bibfield  {author} {\bibinfo {author} {\bibfnamefont {D.}~\bibnamefont
  {Egorova}}, \bibinfo {author} {\bibfnamefont {A.}~\bibnamefont {K{\"u}hl}}, \
  and\ \bibinfo {author} {\bibfnamefont {W.}~\bibnamefont {Domcke}},\
  }\href@noop {} {\bibfield  {journal} {\bibinfo  {journal} {Chem. Phys.}\
  }\textbf {\bibinfo {volume} {268}},\ \bibinfo {pages} {105} (\bibinfo {year}
  {2001})}\BibitemShut {NoStop}%
\bibitem [{\citenamefont {Egorova}\ \emph {et~al.}(2003)\citenamefont
  {Egorova}, \citenamefont {Thoss}, \citenamefont {Domcke},\ and\ \citenamefont
  {Wang}}]{egorova2003modeling}%
  \BibitemOpen
  \bibfield  {author} {\bibinfo {author} {\bibfnamefont {D.}~\bibnamefont
  {Egorova}}, \bibinfo {author} {\bibfnamefont {M.}~\bibnamefont {Thoss}},
  \bibinfo {author} {\bibfnamefont {W.}~\bibnamefont {Domcke}}, \ and\ \bibinfo
  {author} {\bibfnamefont {H.}~\bibnamefont {Wang}},\ }\href@noop {} {\bibfield
   {journal} {\bibinfo  {journal} {J. Chem. Phys.}\ }\textbf {\bibinfo {volume}
  {119}},\ \bibinfo {pages} {2761} (\bibinfo {year} {2003})}\BibitemShut
  {NoStop}%
\bibitem [{\citenamefont {Sachtleben}\ \emph {et~al.}(2017)\citenamefont
  {Sachtleben}, \citenamefont {Mazon},\ and\ \citenamefont {Rego}}]{Kewin}%
  \BibitemOpen
  \bibfield  {author} {\bibinfo {author} {\bibfnamefont {K.}~\bibnamefont
  {Sachtleben}}, \bibinfo {author} {\bibfnamefont {K.~T.}\ \bibnamefont
  {Mazon}}, \ and\ \bibinfo {author} {\bibfnamefont {L.~G.~C.}\ \bibnamefont
  {Rego}},\ }\href {\doibase 10.1103/PhysRevLett.119.090601} {\bibfield
  {journal} {\bibinfo  {journal} {Phys. Rev. Lett.}\ }\textbf {\bibinfo
  {volume} {119}},\ \bibinfo {pages} {090601} (\bibinfo {year}
  {2017})}\BibitemShut {NoStop}%
\bibitem [{\citenamefont {Berendsen}\ \emph {et~al.}(1984)\citenamefont
  {Berendsen}, \citenamefont {Postma}, \citenamefont {van Gunsteren},
  \citenamefont {DiNola},\ and\ \citenamefont {Haak}}]{berendsen1984molecular}%
  \BibitemOpen
  \bibfield  {author} {\bibinfo {author} {\bibfnamefont {H.~J.}\ \bibnamefont
  {Berendsen}}, \bibinfo {author} {\bibfnamefont {J.~v.}\ \bibnamefont
  {Postma}}, \bibinfo {author} {\bibfnamefont {W.~F.}\ \bibnamefont {van
  Gunsteren}}, \bibinfo {author} {\bibfnamefont {A.}~\bibnamefont {DiNola}}, \
  and\ \bibinfo {author} {\bibfnamefont {J.~R.}\ \bibnamefont {Haak}},\
  }\href@noop {} {\bibfield  {journal} {\bibinfo  {journal} {J. Chem. Phys.}\
  }\textbf {\bibinfo {volume} {81}},\ \bibinfo {pages} {3684} (\bibinfo {year}
  {1984})}\BibitemShut {NoStop}%
\bibitem [{\citenamefont {Rego}\ and\ \citenamefont
  {Bortolini}(2019)}]{Rego-JPCC-2019}%
  \BibitemOpen
  \bibfield  {author} {\bibinfo {author} {\bibfnamefont {L.~G.~C.}\
  \bibnamefont {Rego}}\ and\ \bibinfo {author} {\bibfnamefont {G.}~\bibnamefont
  {Bortolini}},\ }\href {\doibase 10.1021/acs.jpcc.8b11057} {\bibfield
  {journal} {\bibinfo  {journal} {J. Phys. Chem. C}\ }\textbf {\bibinfo
  {volume} {123}},\ \bibinfo {pages} {5692} (\bibinfo {year}
  {2019})}\BibitemShut {NoStop}%
\end{thebibliography}%

\pagebreak

\begin{center}
\textbf{\large Supporting Information for \\
Quantum Mechanical Assessment of Optimal Photovoltaic Conditions in Organic Solar Cells}
\end{center}

\setcounter{equation}{0}
\setcounter{figure}{0}
\setcounter{table}{0}
\setcounter{page}{1}
\makeatletter
\renewcommand{\theequation}{S\arabic{equation}}
\renewcommand{\thefigure}{S\arabic{figure}}
\renewcommand{\bibnumfmt}[1]{[S#1]}
\renewcommand{\thepage}{S\arabic{page}}

\section{Semiclassical Electron-Phonon Hamiltonian}

\begin{figure}
    \centering
  \includegraphics[width = 0.75\linewidth]{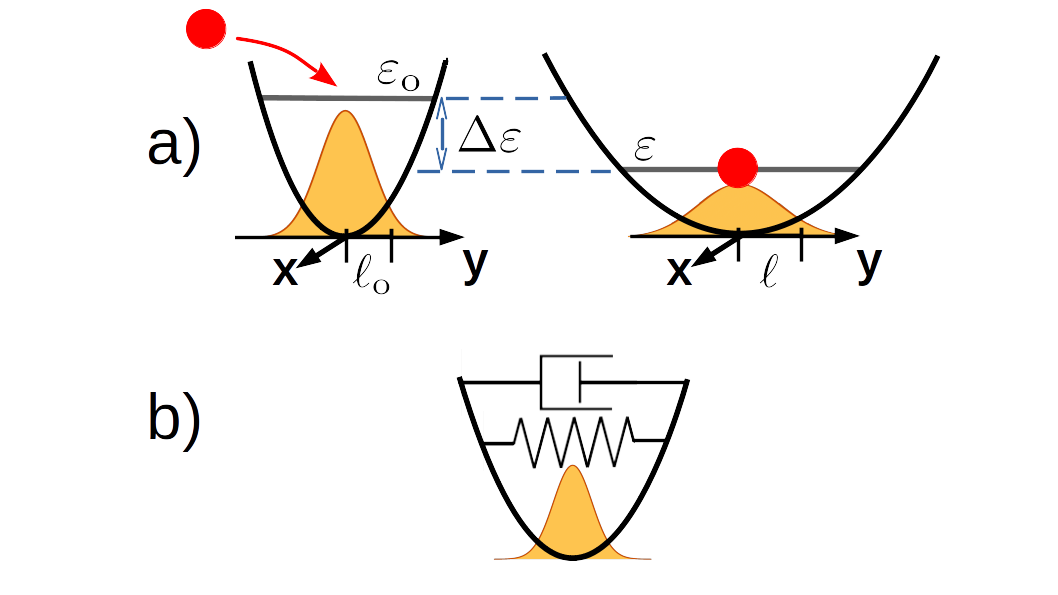}
  \caption{a) Schematics of the electronic relaxation. The parabola describes the confinement potential felt by the electron (or hole) with
its gaussian wavefunction profile. 
On the left-hand  side we  show an empty molecular site, with confinement energy $\varepsilon_\text{o}$ and the corresponding confinement length $\ell_\text{o}$.
On the right-hand  side we  show an occupied molecular site, with on-site energy  $\varepsilon$ and $\ell$. 
The electronic relaxation energy is $\Delta \varepsilon=\varepsilon_\text{o}-\varepsilon$.  
b) The semiclassical electron-phonon system is equivalent to a spring-damper hybrid quantum-classical system.
}
 \label{reorganization} 
\end{figure}

The paraboloids depicted in Figure \ref{reorganization}-a) describe the local confinement potential felt by the electron 
and the hole at a given molecular site. 
We assume that the parabolic confinement potential undergoes a vibrational reorganization whenever charge is transferred in or out of the molecular site.
This mechanism gives rise to coherent vibrational effects that we describe within the framework of the Ehrenfest method.
The density profile of the electron and hole in the molecular site is given by the gaussian
\begin{equation}
g(\vec{r}-\vec{R}) = 
\sqrt{\frac{2}{\pi \ell^2}}\exp\left[-\Big(\frac{\vec{r}-\vec{R}}{\ell}\Big)^2 \right], 
\end{equation}
with $\vec{R}$ designating the position of the molecular site, and  $\ell$ is the confinement radius of the molecular site, given by 
\begin{equation}
\ell = \sqrt{2\hbar^2/(m_e\varepsilon)} 
\label{confine}
\end{equation}
for a gaussian wavepacket of on-site energy $\varepsilon$. 

To write the system Hamiltonian, $H_S$, we assume an orthogonal basis set comprised of diabatic electronic states $\{|i\rangle\}$ associated with the molecular sites,
as described in the paper,
\begin{equation}
H_S \equiv H^{\text{el/hl}} = \sum_{i}^{N}	\left\{E_i^{\text{el/hl}} + \varepsilon_i(t) - \Phi^{\text{el/hl}}_i(t)\right\} |i \rangle \langle i |  
+ \sum_{i \neq j}^{N} V_{ij}(t) |i \rangle \langle j |~.
\label{Hs}
\end{equation} 

The coupling $V_{ij}$ between lattice sites  is given  by 
\begin{equation} 
V_{ij}  = V_0~F(\ell_{i},\ell_{j},d_{ij}) =  
V_0 \left[
 \frac{ 2 \ell_{i}\ell_{j} }{ \ell_{i}^2+\ell_{j}^2} \exp \Big ({ \frac{- d^2_{ij}}{\ell_{i}^2+\ell_{j}^2} } \Big ) 
\right],
\label{hopping}
\end{equation}
where $V_0$ is the bare electronic coupling and the form factor $F$ results from the overlap between
gaussian wavepackets located at lattice sites $i$ and $j$
\begin{equation}
F(\ell_i,\ell_j,d_{ij}) = \int g_i(\vec{r}-\vec{R}_i)~g_j(\vec{r}-\vec{R}_j) dxdy.
\end{equation}
The time dependence of $\varepsilon_i(t)$ and $V_{ij}(t)$, which comprise Eq. (\ref{Hs}),  gives rise to {\it intra}-molecular and {\it inter}-molecular 
vibrational couplings that we associate with the Holstein and Peierls couplings, respectively.

We describe the coherent coupling between the electron-hole states and the fundamental intra-molecular vibrational mode via the 
Ehrenfest method. In this model
the semiclassical electron-phonon system is equivalent to a spring-damper hybrid quantum-classical system, as shown in Figure \ref{reorganization}-b)
The classical Hamiltonian $h_i$ describing  the vibrational mode of site $i$ is given by
\begin{equation}
h_i = \frac{p_i^2}{2\mu} + \frac{\mu \Omega^2}{2}(\ell_i-\ell_{\text{o}})^2 + W^{eh} + Q^{thermo},
\label{intra}
\end{equation}
where $\ell_i$ is the vibrational coordinate associated with the molecular site $i$  and $p_i$ is the conjugate momentum. In addition,
$\ell_{\text{o}}$ is the confinement length for an empty (that is, neutral) site, 
$\mu$ is the effective mass of the vibrational mode and $\Omega$ is the frequency of the relevant normal mode. 
Refer to Table 1 in the paper for a complete list of model parameters.
The energy $W^{eh}$ is the work exchanged with the 
electronic degrees of freedom via the Ehrenfest force $F^{eh}$ and 
$Q^{thermo}$ is the heat exchanged with the classical thermostat. 
A similar hamiltonian  has been considered by Egorova et al.\cite{egorova2001modeling,egorova2003modeling}  for quantum intra-molecular degrees of freedom.
The Ehrenfest force that acts on the vibrational coordinate of a given molecular site $k$ is 
\begin{equation}
F^{eh}_k = F_k^{\text{el}} + F_{k}^{\text{hl}} =  - \frac{\partial U^{\text{el}}}{\partial \ell_k} - \frac{\partial U^{\text{hl}}}{\partial \ell_k}, 
\label{Ehrenfest1}
\end{equation}
with $U^{\text{el}} = Tr[\sigma^{\text{el}}H^{el}]$ and $U^{\text{hl}} = Tr[\sigma^{\text{hl}}H^{hl}]$.
Solving equation (\ref{Ehrenfest1}) we obtain\cite{Artur}
\begin{equation}
F^{eh}_k = - (\sigma^{\text{el}}_{kk} + \sigma^{\text{hl}}_{kk}  ) \frac{d \varepsilon_k}{d \ell_k} 
- \sum_{i \neq j} (\sigma^{\text{el}}_{ij} + \sigma^{\text{hl}}_{ij}) \frac{\partial V_{ij}}{\partial \ell_k}.
\label{Ehrenfest2}
\end{equation}
The first term on the right-hand side (RHS) of Equation (\ref{Ehrenfest2}) is responsible for the relaxation of the molecular site due to its charge occupation; 
this is the classical work exchanged between electronic and vibrational degrees of freedom. 
The second term gives rise to coherent vibronic effects, for it is proportional to
the off-diagonal elements of $\sigma$ and couples the electronic and nuclear coordinates via $\partial V_{ij}/\partial \ell_k$.\cite{Kewin}

The classical equations of motion derived from Eq. (\ref{intra}) are solved with the velocity Verlet algorithm, 
coupled to a classical bath, described by the Berendsen thermostat,\cite{berendsen1984molecular} with relaxation constant  $\gamma^{-1} = 0.1$ ps. 
The perturbing forces change the equilibrium configuration of the parabolic confining potential, which reacts with a restoring force
$F^\text{res} = -\mu\Omega^2(\ell - \ell_{\text{o}})$.

\section{Vibronic Recombination Model}

In order to compare the vibronic recombination model with the three-state recombination model,\cite{Bredas2019,Benduhn2017,Qian2018,JennyNelsonJACS,Chen2021}
let us consider the dynamical equation for recombination, Eq. (6), in the paper
\begin{equation}
\frac{d}{dt}|\Psi^\eta(t)\rangle = -\hat{\Gamma}^{\eta'}_R|\Psi^\eta(t)\rangle~,
\end{equation}
where 
\begin{equation}
\hat{\Gamma}^{\eta'}_R = |\Psi^{\eta'}(t)\rangle~\kappa\hat{F}~\langle \Psi^{\eta'}(t)|~,
\label{rec-oper}
\end{equation}
so that we obtain the following set of coupled equations 
\begin{eqnarray}
\frac{d}{dt}|\Psi^{el}(t)\rangle &=& - \left(\langle\Psi^{hl}(t)|\kappa\hat{F}|\Psi^{el}(t)\rangle\right) |\Psi^{hl}(t)\rangle \\
\frac{d}{dt}|\Psi^{hl}(t)\rangle &=& - \left(\langle\Psi^{el}(t)|\kappa\hat{F}|\Psi^{hl}(t)\rangle\right) |\Psi^{el}(t)\rangle~.
\end{eqnarray}
For the sake of simplifying the argument 
we are assuming here that the electron and hole states are described by wavefunctions. However, in our treatment the electron and hole 
states are described by reduced density matrices because of the system-bath interaction. 
So, let us consider the matrix element $\langle\Psi^{hl}(t)|\kappa\hat{F}|\Psi^{el}(t)\rangle$ that gives rise to the recombination of electrons due to
the operator
\begin{equation}
\kappa\hat{F} = \kappa\sum_i F_{ii}|i\rangle\langle i| +  \kappa\sum_{i\neq j} F_{ij}|i\rangle\langle j|            ~.
\label{F}
\end{equation}
Substituting Eq. \eqref{F} into the previous matrix element and writing the electron and hole wavepackets in the molecular site basis,
$|\Psi^\eta(t)\rangle = \sum_i C^\eta_i(t) |i\rangle$, we get
\begin{eqnarray}
\langle\Psi^{hl}(t)|\kappa\hat{F}|\Psi^{el}(t)\rangle &=& 
\kappa\sum_i \left[C_i^{hl}(t)\right]^*F_{ii}~C_i^{el}(t) +  \kappa\sum_{i\neq j} \left[C_i^{hl}(t)\right]^* F_{ij} ~C_j^{el}(t) \\
&=&
\kappa\sum_i \left[C_i^{hl}(t)\right]^*C_i^{el}(t) +  \kappa\sum_{i} \left(\sum_{j}^{(j\neq i)}\left[C_i^{hl}(t)\right]^* F_{ij} ~C_j^{el}(t)\right),
\label{rec}
\end{eqnarray}
where $F_{ii}=1$ as given by Eq. (5) of the paper.
This expression can be compared with the three-state model.
The first term describes the radiative recombination taking place on molecular site $i$, and $\kappa$ can be associated with the radiative decay constant, 
since it does not depend on temperature or vibrational motion. 
The second term couples distinct molecular sites. For instance, a hole in the donor site $i$ is coupled to electrons in both donor and acceptor sites
via $F_{ij}$, which is the vibronic coupling term that depends on the temperature. This term is a generalization of the LE-CT hybridization considered in 
the three-state model.
However, the choice of a single recombination parameter $\kappa$, associated with the el-hl lifetime, for both contributions is clearly an oversimplification.
Previous studies have treated $\kappa$ differently for radiative and non-radiate recombination processes. 
For instance, Coropceanu and collaborators\cite{Bredas2019,Chen2021} associate $\kappa_r$ with the Einstein spontaneous emission rate for the radiative 
recombination term whereas the Marcus-Levich-Jortner (MLJ) formula was used to define the non-radiative decay rate $\kappa_{nr}$. 
The interplay of charge transfer and non-radiative decay was investigated by atomistic non-adiabatic molecular dynamics simulations in dye-sensitized 
semiconductor interfaces,\cite{Rego-JPCC-2019} where it was revealed a rate competition between the two mechanisms.

Finally, it is worth noting that Eq. \eqref{rec} describes the recombination losses during the exciton diffusion and during the life-time of the 
CT state at the D:A interface.

\newpage

\section{Supporting Simulation Results}

\subsection{Electron-Hole Interaction and Recombination}

\begin{figure}[h]
    \centering
 \includegraphics[width=0.9\linewidth]{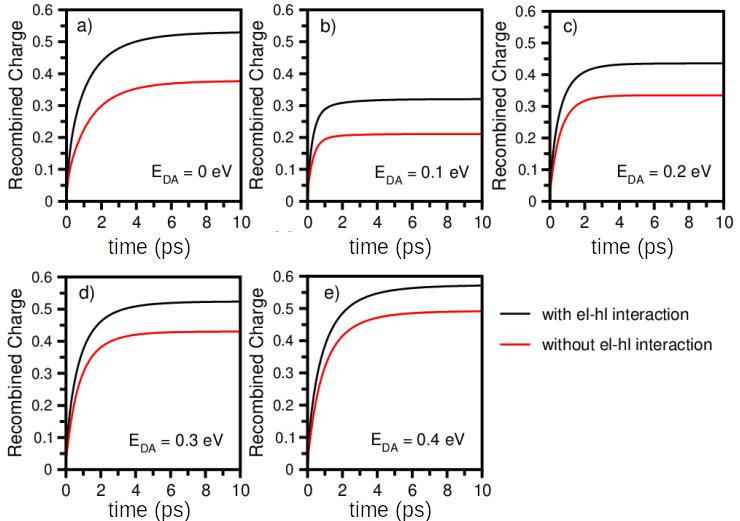}                             
 \caption{Effect of electron-hole interaction $\Phi^{\text{el/hl}}_i(t)|i \rangle \langle i |$, from Equation (4) of the paper, 
on the geminate recombination rate during charge separation. The bath temperature is T$_B$ = 300 K. The energetic driving force $E_{DA}$ is:
0 (a), 0.1 eV (b), 0.2 eV (c), 0.3 eV (d), and 0.4 eV (e).
}
\label{noSOC}
\end{figure}

\newpage

\subsection{Vibrational Dynamics and Recombination}

\begin{figure}[h]
    \centering
 \includegraphics[width=0.8\linewidth]{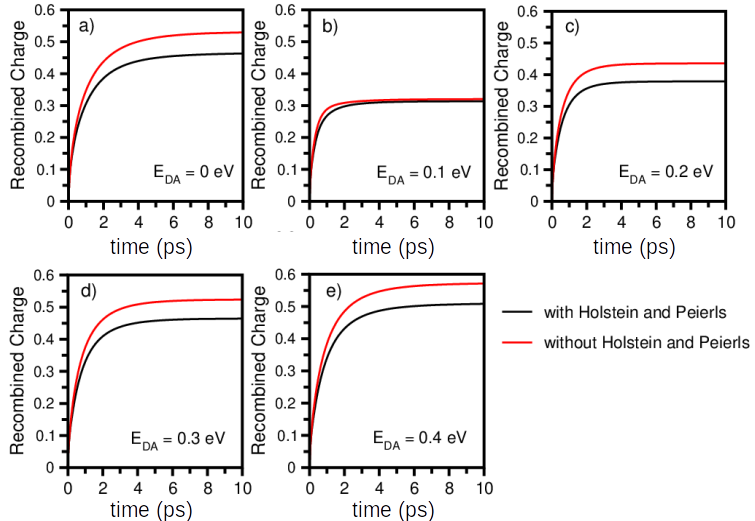}                             
 \caption{Effect of vibrational dynamics (also considered as dynamical disorder) 
on the geminate recombination rate during charge separation. The bath temperature is T$_B$ = 300 K. The energetic driving force $E_{DA}$ is:
0 (a), 0.1 eV (b), 0.2 eV (c), 0.3 eV (d), and 0.4 eV (e).
}                                                                                                          
\label{SOC}
\end{figure}

\clearpage

\subsection{Charge Generation {\it with}  and {\it without} Recombination}

\begin{figure}[htbp]
    \centering
  \includegraphics[width = 0.8\linewidth]{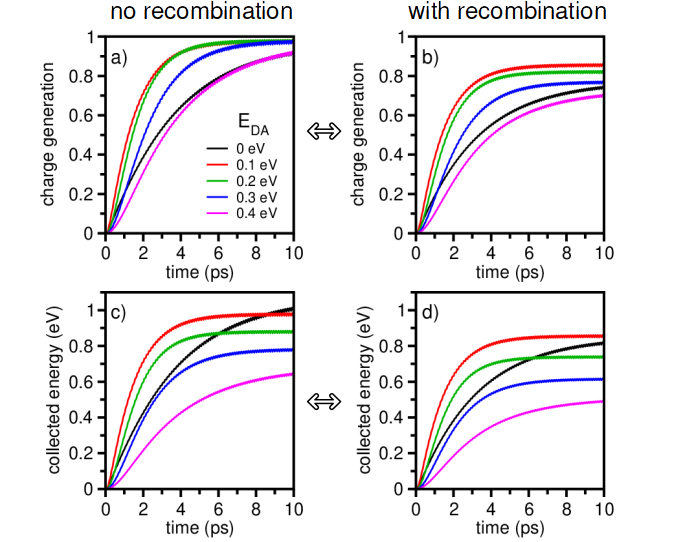}
  \caption{Calculations of time-dependent charge collection, a) and b), and energy collection, c) and d).
Left-hand-side (LHS) panel: {\it no recombination} effects. 
Right-hand-side (RHS) panel: {\it with recombination} effects. 
Calculations performed at T$_B$ = 300 K for various LUMO level offsets, E$_{DA}$, as indicated by the legend.}
 \label{PEC} 
\end{figure}

By comparing the LHS  (no recombination) with the RHS (with recombination) panels of Figure \ref{PEC}, 
we observe that both the charge separation and energy conversion efficiencies decrease as a consequence of recombination for all values of the driving
force $E_{DA}$. For each case, the photovoltaic energy conversion decreased by: 
19\% ($\Delta E^{rec} \approx$ 190 meV) for $E_{DA}$ = 0 (black),
12\%  ($\Delta E^{rec} \approx$ 120 meV) for $E_{DA}$ = 0.1 eV (red),
16\%  ($\Delta E^{rec} \approx$ 140 meV) for $E_{DA}$ = 0.2 eV (green),
21\%  ($\Delta E^{rec} \approx$ 160 meV) for $E_{DA}$ = 0.3 eV (blue) and
24\%  ($\Delta E^{rec} \approx$ 150 meV) for $E_{DA}$ = 0.4 eV (magenta).
The results show that for the activationless regime ($E_{DA}$ = 100 meV $\approx \lambda$) the losses due to recombination are minimized, since charge separation is more
efficient in this regime.\cite{Artur}
As the driving force increases, for $E_{DA}$ = 0.2, 0.3 and 0.4 eV, and the system enters the Marcus inverted regime for charge separation, the losses due to
recombination  increase. 
We ascribe this effect to reduced charge separation, as the driving
force exceeds to the reorganization energy at the interface.
Here the combination of big energetic driving forces and increased recombination renders the overall energy loss for 
$E_{DA}$ = 0.4 eV the highest of all cases, followed by $E_{DA}$ = 0.3 and 0.2 eV, respectively.
However, the biggest energy loss due exclusively to recombination (in absolute value) occurs for $E_{DA}$ = 0, 
because of the much slower electron-hole pair separation rate without the energetic driving
force. Internal electric fields should improve considerably the power conversion efficiency at vanish driving forces.
The simulation results indicate that the condition $E_{DA} \approx \lambda$ has the effect of minimizing the  effects  of the geminate recombination.

\end{document}